\documentclass[10pt,reqno]{amsart}
\usepackage{amsmath,amsxtra,amssymb,amsthm,amsfonts,bm,cite}

\makeatletter
\def\subsection{\@startsection{subsection}{2}%
  \z@{0.5\linespacing\@plus 0\linespacing }{-.5em}%
  {\normalfont\bfseries}}
\makeatother

\usepackage{caption, subcaption}

\usepackage{algorithm,algpseudocode}

\usepackage[cal=cm]{mathalfa}
\usepackage{mathrsfs}

\usepackage{bbold}
\usepackage{bbm}
\usepackage{braket}
\usepackage{dsfont}
\usepackage{mathdots}
\usepackage{mathtools}
\usepackage[shortlabels]{enumitem}
\usepackage{csquotes}
\usepackage{stmaryrd}
\usepackage{graphicx}
\usepackage{stackengine}
\usepackage{scalerel}
\usepackage{array}
\usepackage{makecell}
\newcolumntype{x}[1]{>{\centering\arraybackslash}p{#1}}
\usepackage{tikz}
\usetikzlibrary{shapes.geometric, shapes.misc, positioning, arrows, decorations.pathreplacing, angles, quotes}
\usepackage{booktabs}
\usepackage{xfrac}
\usepackage{siunitx}
\usepackage{centernot}
\usepackage{comment}
\usepackage{chngcntr}

\usepackage{tikz-cd} 

\makeatletter
\def\thmhead@plain#1#2#3{%
  \thmname{#1}\thmnumber{\@ifnotempty{#1}{ }\@upn{#2}}%
  \thmnote{ {\the\thm@notefont#3}}}
\let\thmhead\thmhead@plain
\makeatother

\usepackage{hyperref}
\hypersetup{colorlinks = true,linkcolor = blue,anchorcolor =red,citecolor = blue,filecolor = red,urlcolor = red,
            pdfauthor=author}
\usepackage{cleveref}

\newtheorem{theorem}{Theorem}[section]

\numberwithin{equation}{section}

\theoremstyle{definition}
\newtheorem{definition}[theorem]{Definition}
\newtheorem{remark}[theorem]{Remark}

\newcommand\mycom[2]{\genfrac{}{}{0pt}{}{#1}{#2}}

\newcommand{\norm}[1]{\lVert #1 \rVert}

\newcommand{\mc}[1]{\mathcal{#1}}

\newcommand{\w}[1]{\widetilde{#1}}

\def\q {\quad}
\def \qq {\qquad}
\def \l{\langle}
\def \r{\rangle}
\def \h{\hat}

\def\bb{\begin{equation}
  \left\{\ 
   \begin{aligned} }
\def\ee{   \end{aligned}
  \right.
  \end{equation}}

\def\mm{ \left[
 \begin{matrix}}
\def\nn{\end{matrix} \right] } 

\def \lad {\lambda}
\def \d{\delta}
\def \ep {\varepsilon}
\def \ww {\omega}

\def \R{\mathbb{R}}
\def \C{\mathbb{C}}

\def \ff{\textit{\textbf{F}}}

\def \dd {\cdot}
\def \t {\times}

\def \si{\sigma}

\def \vp {\varphi}
\def \vr{\varrho}

\def \up {\uparrow}
\def \down {\downarrow}

\DeclareMathOperator{\tr}{Tr}

 \def \sss   {\scriptscriptstyle}

\title[Variational embedding for ground-state energy problems]{Quantum variational embedding for  ground-state energy problems: sum of squares and cluster selection}

\begin{document}

\author[B. Li]{Bowen Li} %
\address[B. Li]{Department of Mathematics, Duke University, Durham, NC 27708.}
\email{bowen.li200@duke.edu}

\author[J. Lu]{Jianfeng Lu}
\address[J. Lu]{Departments of Mathematics, Physics, and Chemistry, Duke University, Durham, NC 27708.}
\email{jianfeng@math.duke.edu}

\begin{abstract}
We introduce a sum-of-squares SDP hierarchy 
approximating the ground-state energy from below for quantum many-body problems, with a natural quantum embedding interpretation. We establish the connections between our approach and other variational methods for lower bounds, including the variational embedding, the RDM method in quantum chemistry, and the Anderson bounds. 
Additionally, inspired by the quantum information theory, 
we propose efficient strategies for optimizing cluster selection to tighten SDP relaxations while staying within a computational budget. Numerical experiments are presented to demonstrate the effectiveness of our strategy. As a byproduct of our investigation, we find that quantum entanglement has the potential to capture the underlying graph of the many-body Hamiltonian.
\end{abstract}

\maketitle
\section{Introduction}
Computing the ground-state energy of a strongly correlated many-body system is a notoriously challenging
problem in quantum physics and chemistry, due to the exponential scaling of the problem size in the particle number. This curse of dimensionality makes direct solutions for large-scale systems impractical and highlights the need to develop numerical techniques that can accurately approximate the problem at an affordable computational cost.

Quantum embedding theory is a powerful set of computational techniques for simulating quantum many-body systems. Its general idea is to exploit the locality of particle interactions to
divide the large, complex system into smaller, more manageable subsystems, known as fragments, that can be solved with high accuracy in a consistent way \cite{lindsey2019quantum,sun2016quantum}. One such approach is density functional embedding theory (DFET), which partitions the total system into a fragment and an environment, using an effective embedding potential to account for the environment's influence on the subsystem \cite{cortona1991self,manby2012simple,wesolowski1993frozen}. Other prominent examples include the dynamical mean-field theory (DMFT) \cite{georges1996dynamical,kotliar2006electronic,held2007electronic} 
and the density-matrix-embedding theory (DMET) \cite{knizia2012density,wouters2016practical,knizia2013density}, both of which map the full interacting system to an impurity embedded in a non-interacting bath. A known challenge for quantum embedding theory is how to combine the high-level calculation on the fragment with the low-level calculation on the environment. 

Another class of methods for solving the ground-state energy problem relies on the variational principle: 
$E_0[H] = \{\bra{\vp}H\ket{\vp};\, \ket{\vp}\in \mc{H},\, \l \vp|\vp \r = 1 \}$, where $\ket{\vp}$ denotes the wave function. Since a 
typical many-body Hamiltonian $H$ involves only local interactions, 
the degree of freedom of $H$ scales polynomially in the number of particles or orbitals. Moreover, the possible wave function that could be a ground state is not arbitrary and possesses special characteristics such as low entanglement. 
Therefore, it is promising to consider the optimization over 
a restricted class of wave functions that can be parameterized efficiently and 
capture the main properties of the ground states, which necessarily provide the upper bounds to the exact energy $E_0[H]$.
Many successful algorithms are based on this fundamental principle, for instance, the 
density-matrix renormalization group (DMRG) algorithm \cite{white1992density,schollwock2005density} and, more generally, the tensor network methods \cite{verstraete2008matrix,verstraete2004renormalization}.

In addition, for a local Hamiltonian $H$, the energy $\bra{\vp}H\ket{\vp}$ of a wave function $\ket{\vp}$ is essentially 
determined by the correlations or the reduced density matrices (RDMs) of the state $|\vp\r\l\vp|$. This fact has been utilized in method development since the 1950s.  
Mayer \cite{mayer1955electron} proposed to compute the energy of an electronic (fermionic) system in terms of 1-RDM and 2-RDM, but meanwhile one has to characterize the conditions under which the RDMs are consistent 
with a global wave function \cite{coleman1963structure}. Such \emph{representability} problem is fundamental and
has been studied extensively in past decades 
\cite{walter2014multipartite,mazziotti2012structure,gidofalvi2004boson,klyachko2004quantum}. 
It was recently proven that the representability problem and the related ground-state energy problem are QMA-hard  \cite{oliveira2005complexity,liu2007quantum,liu2006consistency,cubitt2016complexity}, making them difficult to solve 
even for quantum computers. Despite this, many efforts have been devoted to finding the approximate representability 
conditions, resulting in the development of the variational RDM methods that approximate the energy 
from below by semidefinite programming (SDP) problems; see \cite{zhao2004reduced,fukuda2007large,mazziotti2005variational,mazziotti2006variational,mazziotti2002variational,mazziotti2004realization,barthel2012solving,baumgratz2012lower,haim2020variational} and references therein.  

Recently, a quantum variational embedding method also formulated as an SDP relaxation to $E_0[H]$ was introduced by Lin and Lindsey \cite{lin2022variational}. 
This method employs quantum marginals as optimization variables with local consistency constraints and a global semidefinite constraint generalized from \cite{khoo2020semidefinite}. Since it does not involve any model approximation error, the relaxation given by the variational embedding is guaranteed to be tighter than the one in \cite{khoo2020semidefinite} which is
based on the strictly correlated
electron limit. However, the relationship between the variational embedding and the widely used 2-RDM method, or equivalently, the Lasserre (sum-of-squares) hierarchy \cite{lasserre2001global,pironio2010convergent}, has not been fully explored yet. Additionally, the embedding formulation suffers from the exponential scaling with the cluster size. 
These questions and concerns motivate the current work.

\subsection*{Contribution and related works} The sum-of-squares (SOS) hierarchy is a  popular technique in  polynomial optimization and computational complexity theory \cite{kothari2017sum,lasserre2001global,pironio2010convergent} with fruitful applications in various fields, such as the Goemans-Williamson algorithm for the 
MAX-CUT problem and the RDM method reviewed above. 
Noting the equivalence between the SOS hierarchy and the RDM method \cite[Section 5]{pironio2010convergent}, in the present work, we shall generalize the variational embedding from the perspective of sum-of-squares.  Furthermore, to tighten the embedding schemes with given computational resources, we will propose some simple yet effective strategies to optimally select the clusters.

Our results apply to all particle statistics (spins, fermions, and bosons), although we focus our discussion on quantum spin and fermionic systems for simplicity. We first 
establish the connections between the ground-state energy problem and the quantum multi-marginal optimal transport (OT) as in \cite{feliciangeli2023non}. While this step is not really necessary for the variational embedding, it provides a more comprehensive set of primal-dual formulations for the SDP relaxations and enables further algorithm development based on entropic regularization \cite{lindsey2023fast,feliciangeli2023non}. Then we introduce a variant of the standard SOS hierarchy to relax the positivity constraint in the dual formulation, resulting in the SDP problems $E_0^{ (k)}$  \eqref{eq:kthconst}. 
We show that the relaxations $E_0^{ (k)}$ 
can be formulated as an optimization of $\sum_{\gamma < \d} E_0[\h{H}_{\gamma \d}^{{\rm eff}}]$ over the effective Hamiltonians 
 defined on fragments $V_{\gamma \d}$ and coupled with each other, which implies a quantum embedding interpretation and a connection with Anderson bounds 
 \cite{anderson1951limits,wittmann1993bounds}, as shown in equations \eqref{scheme_lag_2}, \eqref{auxclaim2}, and Remark \ref{rem:inter_eff}. We elaborate on how 
  the recently proposed variational embedding formulation \cite{lin2022variational} fits into our SOS relaxation hierarchy and demonstrate that the lowest-order embedding scheme is tighter than the RDM method with the standard $P,Q,G$ conditions. We also discuss
 how to exploit RDM conditions to tighten the variational embedding. 
 
 The exponential scaling in cluster size is a fundamental challenge for embedding techniques. In this study, we investigate the potential for improving the accuracy of the SDP approximation by optimally selecting the clusters with a given cluster size. Similar ideas have been considered in many algorithms for electronic structure calculations. To name a few, Legeza et al. \cite{legeza2003optimizing} 
studied the correlation and entanglement of the target state to suggest an optimal ordering of the sites for the 
DMRG method, which leads to faster convergence.  Krumnow et al. \cite{krumnow2016fermionic} used the Gaussian mode transformations to optimize fermionic orbitals for the 
tensor network ansatz. Li et al. \cite{li2020optimal} 
employed partial unitary matrices to compress orbitals and improve the full configuration 
interaction method.

Recall the basic fact about the quantum embedding theory: it is highly accurate at the local scale while less accurate globally, which makes it very effective when the Hamiltonian is non-interacting, or the ground state is a product state. We hence expect that the clusters aligning with the correlation or entanglement structure of the ground state can enhance the capability of the variational embedding. Namely, the ground state is strongly correlated inside the chosen clusters while weakly correlated between them. We will see that it is indeed the case. Letting $\{C_i\}$ be given clusters, we find that if the mutual information of the marginal of the ground state on $C_i \cup C_j$ is larger, then grouping $C_i$ and $C_j$ can yield a tighter variational embedding. Building on this observation, we propose several strategies by greedy algorithms to gradually select optimal clusters with large correlations. The numerical results demonstrate that our cluster-selection strategy can tighten the SDP relaxations with significant error reductions 
with high probability. It is also interesting to note that quantum entanglement can accurately recover the underlying graph structure of the Hamiltonian in some cases.

\subsection*{Organization}
Section \ref{sec:basic} gives the preliminaries of the fermionic 
ground-state energy problem and the relations with the multi-marginal quantum OT. In section \ref{sec:sosemb}, we propose a sum-of-squares SDP hierarchy to approximate the 
ground-state energy that generalizes the variational embedding method, and connect it with other existing relaxation techniques. The extension to the quantum spin case will also be discussed. 
Section \ref{sec:multiembedding}
considers the optimal cluster selection based on mutual information. In Section \ref{sec:numerical}, we test the variational embedding with optimized clusters for various quantum many-body problems with randomly generated graphs or coefficients.

\section{Ground-state energy and multi-marginal quantum optimal transport} \label{sec:basic}

In this section, we will introduce the second quantization formulation of the fermionic ground-state energy problem and its connection with multi-marginal quantum optimal transport. 

Let us first fix some notation used throughout this work. We write $[n]$ for the set $\{1,2,\ldots, n\}$ with an integer $n$. For a Hilbert space $\mc{H}$, we denote by $\mc{H}'$ its dual space of continuous linear functionals on $\mc{H}$. The Hermitian adjoint of a bounded linear operator $A$ on $\mc{H}$ 
is denoted by $A^\dag$. Moreover, for any matrix $Y$, its conjugate matrix $\overline{Y}$ is defined by $\overline{Y}_{ij} = \overline{Y_{ij}} $ (its conjugate transpose is given by $Y^\dag$ as above).

\subsection{Fermionic systems} \label{sec:fermboso} We consider a many-fermion system with indistinguishable particles, where a wave function is the linear combination of antisymmetric tensor products of single-particle states. In the second quantization \cite{negele2018quantum,coleman2015introduction}, the basis states are described by the occupation numbers. Let $d > 0$ be the number of possible orbitals of the fermions. The \emph{occupation number basis} for the system is defined as $\ket{{\bf n}} = \ket{n_1, n_2, \ldots, n_d}$, with $n_i \in \{0,1\}$ 
indicating whether or not the $i$th orbital is occupied by a fermion, by the Pauli exclusion principle.
Note that this system contains at most $d$ fermions. We define the \emph{Fock space} $\mc{F}_{d}$ by the vector space spanned by all the basis states $\ket{{\bf n}}$ with an inner product such that $\ket{{\bf n}}$ are orthonormal. It is easy to see that $\mc{F}_d$ is isomorphic to the 
multi-qubit space $\C^{2^d} \simeq \otimes_{i = 1}^d \C^{2}$. Recall the 
anti-commutator $\{A,B\} = AB + BA$ for two operators $A, B$. 
We define the \emph{annihilation operators} $a_i$ by
\begin{align*}
     a_i \ket{{\bf n}} =   (-1)^{\sum_{j < i}n_j} n_i\ket{n_1,\ldots,n_{i-1}, 1- n_i, n_{i+1}, \ldots, n_d},
\end{align*}
and the \emph{creation operators} $a_i^\dag$ by their adjoints,   
which satisfy the canonical anti-commutation relations (CARs): 
\begin{equation} \label{eq:commu}
    \{a_i, a_j\} = \{a_i^\dagger, a_j^\dagger\} = 0\,, \quad \{a_i,a_j^\dagger\} = \delta_{ij}\,.
\end{equation}
We write $\ket{\Omega}$ for the \emph{vacuum state} $\ket{0,\ldots,0}$, and then have
$
    \ket{n_1, \ldots, n_d} = (a_1^\dag)^{n_1} \ldots  (a_d^\dag)^{n_d} |\Omega \r
$. 

Now let $\mc{A}_d$ be the unital $*$-algebra generated by the operators $1$ and 
$a_i^\dag, a_i$ for $i \in [d]$ with the CARs \eqref{eq:commu}, known as the Clifford algebra ${\rm Cliff}(2d, \C)$ \cite{woit2017quantum} and denoted as follows, 
\begin{align} \label{cliff_alg}
    \mc{A}_d: = \left\l \{1\} \cup \{a_i\,;\ i \in [d]\} \right\r,
\end{align}
whose linear basis of $\mc{A}_d$ is given by 
\begin{align} \label{def:basis_cli}
\vp_1\vp_2 \cdots \vp_d\,,\q \vp_i \in \{a_i^\dag a_i, a_i a_i^\dag, a_i, a_i^\dag\}\,.
\end{align}
We always use the angle 
brackets to denote the $*$-algebras generated as in \eqref{cliff_alg}.
Noting that an element in $\mc{A}_d$ is a polynomial in operators $a_i$ and $a_i^\dag$, by CARs \eqref{eq:commu} we can assume that it is normally ordered, that is, all the creation operators are to the left of all the annihilation operators.
The fermionic \emph{Hamiltonians} $H$ are defined as elements in $\mc{A}_d^{{\rm sa}}: = \{A \in \mc{A}_d\,;\ A^\dag = A\}$, the subset of self-adjoint elements. We also introduce the \emph{number operator} $\hat{n}_{i}:=a_{i}^{\dagger}a_{i}$ counting the number of particles in the orbital $i \in [d]$, and the total number operator $\hat{N}:=\sum_{i=1}^{d}\hat{n}_{i}: \mc{F}_d \to \mc{F}_d$. It is clear that $\h{N}$ is
diagonalizable with eigenvalues $\{0,1,\ldots,d\}$, and the associated $N$-eigenspace of $\h{N}$ consists of all the states of $N$ fermions. 

Consider a Hamiltonian $H$ commuting with $\h{N}$ and hence preserving the particle number. The \emph{ground-state energy} problem constrained in the $N$-particle space reads as follows:
\begin{align} \label{def:ground}
    E_{0,N}[H] = \inf_{\ket{\vp} \in \mc{F}_{d}}\left\{\bra{\vp} H \ket{\vp}\,;\ \braket{\vp\, |\, \vp} = 1\,,\  \bra{\vp} \h{N} \ket{\vp} = N  \right\}.
\end{align}
Recall that a linear functional $\vr: \mc{A}_{d}  \to \C$ is positive if $\vr(h) \ge 0$ for any $h \ge 0$, and a positive functional is also Hermitian-preserving, i.e., $\vr(h^\dag) = \vr(h)^\dag$. We define the \emph{quantum states} $\vr \in \mc{A}_d'$ as the normalized positive linear maps on the algebra $\mc{A}_{d}$, i.e., $\vr \ge 0$ with
$\vr(1) = 1$, and we denote by $\mc{D}(\mc{A}_{d})$ the convex set of all quantum states.
With these notions, we reformulate \eqref{def:ground} as 
\begin{equation} \label{eq:ferm}
E_{0,N}[H] = \inf_{\vr \in \mc{D}(\mc{A}_{d})} \left\{\vr(H)\,;\  \vr(\h{N}) = N \right\}\,,
\end{equation}
where
$\vr(H)$ can be viewed as the quantum expectation of $H$ with respect to the state $\vr$.
 By introducing the Lagrange multiplier $\mu$ (called the \emph{chemical potential}) for the constraint $\vr(\h{N}) = N$, we obtain, from Sion's minimax theorem \cite{komiya1988elementary}, 
\begin{align} \label{model_fermiboson}
E_{0,N}[H] = \sup_{\mu \in \R} \inf_{\vr \in \mc{D}(\mc{A}_{d})} \vr(H - \mu \h{N}) + \mu N\,.
\end{align}
In practice, the chemical potential $\mu \in \R$ may be fixed in advance based on the Hamiltonian, or adjusted during the numerical computation, such as in the DMFT, to ensure the specified total particle number \cite{lindsey2019quantum,jia2017robust,lin2019mathematical}.
Without loss of generality, we absorb the term $\mu \h{N}$ into $H$ and focus on the following eigenvalue problem:
\begin{align} \label{subtarget}
    E_{0}[H] = \inf_{\vr \in \mc{D}(\mc{A}_{d})} \vr(H)\,.
\end{align}
 Once the problem \eqref{subtarget} is solved,  the $N$-particle ground-state energy $E_{0,N}[H]$ in \eqref{eq:ferm} can be easily computed by an additional one-dimensional 
 optimization problem in $\mu \in \R$.

\subsection{Connection with multi-marginal quantum OT}
\label{sec:mopfer} 
We consider the disjoint partition of orbitals $[d]$:
\begin{align}\label{eq:index_cluster_fermi}
    [d] = \bigcup_{j = 1}^{d_c} C_j\,, \q d_c \le d\,,
\end{align}
and write $C_{ij}:= C_i \cup C_j$. For any cluster $C \subset [d]$, 
we define the associated algebra 
$\mc{A}_{C}: = \l \{1\} \cup \{a_i\,;\ i \in C\}\r$, and denote by $\mc{D}(\mc{A}_C)$ the set of quantum states on $\mc{A}_C$. Moreover, we define the marginal $\vr^{C} \in \mc{D}(\mc{A}_C)$ of a state $\vr \in \mc{D}(\mc{A}_d)$ by its restriction to $\mc{A}_{C}$, i.e., $\vr^{C} := \vr|_{\mc{A}_{C}}$. Let $H \in \mc{A}^{{\rm sa}}_{d}$ be a fermionic Hamiltonian with the following decomposition for the partition \eqref{eq:index_cluster_fermi}:
\begin{align} \label{hamiltonian_decom}
    H = \sum_{1 \le i \le d_c} H_{C_i} + \sum_{1 \le i < j \le d_c} H_{C_{ij}}\,,
\end{align}
where $H_{C_i} \in \mc{A}^{{\rm sa}}_{C_i}$ and $H_{C_{ij}}\in \mc{A}^{{\rm sa}}_{C_{ij}}$ are local Hamiltonians. 
We shall interpret the ground-state energy \eqref{subtarget} as a quantum multi-marginal
OT. Note similar results have been obtained in the finite temperature regime in a very recent work \cite{feliciangeli2023non}.

Assuming that the marginals 
 $\vr^{C_i} \in \mc{D}(\mc{A}_{C_i})$ are known, 
minimizing the energy of $H$ conditional on $\vr^{C_i}$ can be expressed as
 \begin{align} \label{eq:condition_energy_fermi}
          E_0[H](\vr^{C_1},\ldots,\vr^{C_{d_c}}) = \inf\Big\{\vr(H)\,;\ \vr \in \Gamma(\vr^{C_1},\ldots,\vr^{C_{d_c}})\Big\}\,,
\end{align}
where $\Gamma(\vr^{C_1},\ldots,\vr^{C_{d_c}})$ is the set of coupling quantum states:
\begin{align} \label{def:couple}
    \Gamma(\vr^{C_1},\ldots,\vr^{C_{d_c}}) := \left\{\vr \in \mc{D}(\mc{A}_d)\,;\ \vr|_{\mc{A}_{C_j}}  = \vr^{C_j},\  j \in [d_c]\right\}.
\end{align}
Further minimizing the marginals $\vr^{C_i}$ recovers the ground-state energy:
\begin{equation} \label{eq:con_ground_fermi}
    E_0[H] = \inf \Big\{E_0[H](\vr^{C_1},\ldots,\vr^{C_n})\,;\ \vr^{C_i}\in \mc{D}(\mc{A}_{C_i})\Big\}.
\end{equation}
It is natural to regard the problem \eqref{eq:condition_energy_fermi} as a quantum generalization of the classical multi-marginal OT, which admits the following dual formulation: 
\begin{align} \label{model_dual_fermi}
 E_0[H](\vr^{C_1},\ldots,\vr^{C_{d_c}}) = \sup_{Y_{C_i} \in \mc{A}^{{\rm sa}}_{C_i}} \bigg\{\sum_i \vr^{C_i}(Y_{C_i})\,;\  H - \sum_{i \in [d_c]} Y_{C_i} \ge 0 \bigg\}\,,
\end{align}
where $Y_{\sss C_i}$ is the multiplier (quantum Kantorovich potential) for the constraint $\vr|_{\mc{A}^{{\rm sa}}_{\sss C_i}} = \vr^{\sss C_i}$. We remark that the existence of the maximizer for \eqref{model_dual_fermi} is guaranteed for the case $\vr^{\sss C_i} > 0$ by the standard SDP theory \cite{vandenberghe1996semidefinite}, while 
the general case $\vr^{\sss C_i} \ge 0$ requires additional assumptions for the Hamiltonian $H$; see \cite[Appendix B.3] {cole2021quantum} for a counterexample.
We introduce $X_{\sss C_i} := Y_{\sss C_i} - H_{\sss C_i}$ and rewrite \eqref{model_dual_fermi} as
\begin{align*}
   E_0[H](\vr^{C_1},\ldots,\vr^{C_{d_c}}) = \sup_{X_{C_i} \in \mc{A}^{{\rm sa}}_{C_i}} \bigg\{\sum_i \vr^{C_i}(X_{C_i} + H_{C_i}) \,;\  \sum_{1 \le i < j \le d_c} H_{C_{ij}} - \sum_{i \in [d_c]} X_{C_i} \ge 0\bigg\}\,,
\end{align*}
which implies, by \eqref{eq:con_ground_fermi} and again the Sion’s minimax theorem,
\begin{align} \label{eq:dual_ground}
     E_0[H] = \sup_{X_{C_i} \in \mc{A}^{{\rm sa}}_{C_i}} \bigg\{\sum_i E_0\left[X_{C_i} + H_{C_i}\right] \,;\  \sum_{1 \le i < j \le d_c} H_{C_{ij}} - \sum_{i \in [d_c]} X_{C_i} \ge 0 \bigg\}\,.
\end{align}
Here $E_0\left[X_{\sss C_i} + H_{\sss C_i}\right]$ denotes the ground-state energy of the Hamiltonian $X_{\sss C_i} + H_{\sss C_i}$ as in \eqref{subtarget}. 

The decomposition \eqref{hamiltonian_decom} is motivated by the Hubbard model, which is possibly the simplest model of interacting fermions but with rich phase transitions and correlation phenomena, and has served as the paradigmatic model for the study of the strongly correlated fermionic system \cite{lieb2004hubbard,qin2022hubbard,arovas2022hubbard}. In our experiments, we will consider the $t$--$U$ spinless Hubbard model:
\begin{align} \label{eq:hamiltonian_spinless}
     H = - t\sum_{i \sim j} \left(a_i^\dag a_j +  a_j^\dag a_i \right) + \sum_{i \sim j} U \Big(\h{n}_i - \frac{1}{2}\Big)\Big(\h{n}_j - \frac{1}{2}\Big)\,,
\end{align}
where the first term is the kinetic energy of the system with the hopping integral $t \ge 0$; the second term is the on-site interaction with strength $U \in \R$ ($U \ge 0$ and $U \le 0$ are for the cases of repulsive and attractive fermions, respectively); $i \sim j$ denotes the adjacency of the underlying graph of $H$ with vertices $[d]$. 

One may also consider the spinful Hubbard model given by
\begin{align}  \label{eq:hamiltonian_spin}
    H = - t \sum_{i \sim j} \sum_{\si = \uparrow, \downarrow} \left(a_{i\sigma}^\dag a_{j \sigma} +  a_{j \sigma}^\dag a_{i \sigma} \right)  + U \sum_i \h{n}_{i\uparrow} \h{n}_{i\downarrow}\,,
\end{align}
where each orbital has two states: spin up and spin down, indexed by $(i,\si)$ with $i \in [d]$ and $\si = \uparrow, \downarrow$, and other notations have the same meanings as in \eqref{eq:hamiltonian_spinless}. Note that the orbital-spin index $(i,\si)$ can be mapped to the single index by defining $b_{2i - 1} = a_{i\uparrow}$ and $b_{2i} = a_{i\downarrow}$ for $i \in [d]$, and that for any cluster partition \eqref{eq:index_cluster_fermi}, the Hubbard models
\eqref{eq:hamiltonian_spinless} and \eqref{eq:hamiltonian_spin} can be formulated as  \eqref{hamiltonian_decom}.

It should be emphasized that the form of the Hamiltonian \eqref{hamiltonian_decom} is not essential for our discussions. All the results presented in this work can be easily adapted for  general local Hamiltonian:
\begin{align} \label{generalham}
      H = \sum_{1 \le i \le d_c} H_{C_i} + H_{{\rm int}}\,, 
\end{align}
where $H_{{\rm int}} \in \mc{A}_d^{{\rm sa}}$ denotes the interaction between clusters. For example, we can consider Hamiltonians arising from electronic structure problems:
\begin{align}\label{eq:hamiltonian_general}
    H = \sum_{ij} t_{ij}a_i^\dag a_j + \sum_{ijkl}v_{ijkl}a_i^\dag a_j^\dag a_l a_k\,,
\end{align}
where the coefficients $t_{ij}$ and $v_{ijkl}$ are obtained from the molecular Hamiltonian and the single-particle orbital basis functions  \cite{lin2019mathematical}.

\section{Variational embedding via sum-of-squares hierarchy} \label{sec:sosemb}

In Section \ref{subsec:SOS}, we introduce a variant of sum-of-squares SDP hierarchy approximating $E_0[H]$ from below that can be interpreted as a quantum embedding. Some practically efficient embedding schemes will be derived in Section \ref{subsec:embedding}. Then we relate our approach to existing ones in Section \ref{subsec:connection}. The extension to the quantum spin system will be discussed in Section \ref{subsec:spin}.

\subsection{Sum-of-squares and hierarchical relaxations} \label{subsec:SOS}
\begin{figure}[!htbp]
  \centering
  \includegraphics[width=0.25\textwidth]{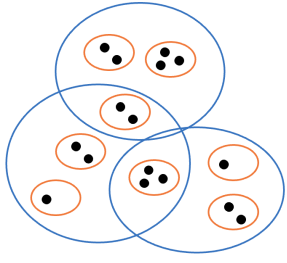}
  \caption{An example of 16 orbitals (black dots) with 8 clusters (orange  circles) in \eqref{eq:index_cluster_fermi} and 3 groups (blue circles) in \eqref{eq:group_v}.}
  \label{fig:group_cluster}
\end{figure}

To make the discussion as general as possible, we associate the Hamiltonian \eqref{hamiltonian_decom} with a graph $(V,E)$ with  vertices $V := [d_c]$ and edges $E := \{i \sim j\,;\ H_{C_{ij}} \neq 0\}$, and consider the grouping of nodes with $M = O({\rm poly}(d))$ being the number of groups: 
\begin{align} \label{eq:group_v}
V = \bigcup_{\gamma \in [M]} V_\gamma\,, 
\end{align}
where $V_\gamma$ may overlap with each other; see Figure \ref{fig:group_cluster} below for an illustration example of clusters of orbitals \eqref{eq:index_cluster_fermi} and their grouping \eqref{eq:group_v}.
For convenience, we often identify the set $V_\gamma$ and 
the associated subset of orbitals $\{i\in C_{j}\,;\ j \in V_{\gamma}\}$, and use the terms \emph{group} and \emph{cluster} for $V_\gamma$ interchangeably. 
We also define $V_{\gamma \delta}: = V_\gamma \cup V_\delta$ and the associated algebra $\mc{A}_{V_{\gamma \delta}} := \big\l \{1\}\cup\{a_{i}\,;\,i\in C_{j},\ j \in V_{\gamma \delta}\} \big \r$. Note that both $\{V_\gamma\}$ and $\{V_{\gamma \d}\}$ form partitions of orbitals, and there always holds 
\begin{align} \label{eq:partitionham}
    \sum_{1 \le i < j \le d_c} H_{C_{ij}} - \sum_{i \in [d_c]} X_{C_i} = \sum_{\gamma < \d} H_{\gamma \delta}(\{X_{C_i}\})\,,
\end{align}
for some local interacting Hamiltonians $H_{\gamma \d} \in \mc{A}^{{\rm sa}}_{V_{\gamma \d}}$ on $V_{\gamma \d}$ depending on $\{X_{C_i}\}$. It is helpful to
derive the general representation for $\{H_{\gamma \d}\}_{\gamma < \d}$. We define the edge sets $E_{\gamma \delta}: =\{i \sim j\,;\ i,j \in V_{\gamma \d} \}$
connecting the nodes in $V_{\gamma \d}$, which
satisfy $E = \cup_{\gamma < \d} E_{\gamma \d}$, and we introduce 
\begin{align*}
I_{i \sim j}:= \{(\gamma, \delta)\,;\ i \sim j \in E_{\gamma \delta}\,,\ \gamma < \d \} \q \text{and} \q  J_i : =  \{j \in [d_c]\,; \ i \sim j \in E_{\gamma \delta}\}\,. 
\end{align*}
With these notations, we then have
\begin{align} \label{eq:split_hamil}
    H_{\gamma \delta}(\{X_{C_i}\}) = \sum_{i \sim j \in E_{\gamma\delta}} w_{ij}^{\gamma\delta} H_{C_{ij}} - \left( w_{i \sim j,i}^{\gamma \delta} X_{C_i} + w_{i \sim j,j}^{\gamma \delta} X_{C_j} \right)\,,
\end{align}
where $w_{ij}^{\gamma\delta} \ge 0$ and $w_{i \sim j,i}^{\gamma \delta} = w_{j \sim i,i}^{\gamma \delta} \ge 0$ are the weights 
satisfying, for each $i \sim j \in E$, 
\begin{align*}
    \sum_{(\gamma, \delta) \in I_{i \sim j}} w_{ij}^{\gamma \delta} = 1 \q \text{and}\ \q \sum_{j \in J_i}  \sum_{(\gamma, \delta) \in I_{i \sim j}}  w_{i \sim j,i}^{\gamma \delta} = 1\,.
\end{align*}
In what follows, we usually omit the argument $\{X_{C_i}\}$ of $H_{\gamma \delta}(\{X_{C_i}\})$ for simplicity.

In view of \eqref{eq:dual_ground} and \eqref{eq:partitionham}, we consider the following convex cone of 2-cluster fermionic Hamiltonians in the product vector space $\prod_{\gamma < \d} \mc{A}_{V_{\gamma \d}}$:
\begin{align} \label{eq:coneham}
    \mc{L}_+ :=  \Big\{\{h_{\gamma \d}\}_{\gamma < \d}   \,;\ h = \sum_{\gamma < \d}h_{\gamma \d} \ge 0\,,\ h_{\gamma \d}\in \mc{A}_{V_{\gamma \d}}  \Big\}\,.
\end{align}
Then the ground-state energy problem \eqref{subtarget} is a non-commutative polynomial optimization over the Clifford algebra with a sparse positive polynomial constraint:
\begin{align} \label{eq:dual_ground2}
     E_0[H] = \sup_{X_{C_i} \in \mc{A}^{{\rm sa}}_{C_i}} \Big\{\sum_i E_0\left[X_{C_i} + H_{C_i}\right] \,;\  \{H_{\gamma \d}\}_{\gamma < \d} \in \mc{L}_+ \Big\}\,.
\end{align}
Here the sparsity means that each monomial in $\sum_{\gamma < \d} H_{\gamma \d}$ depends on at most one cluster $V_{\gamma \d}$ for some $\gamma < \d$. It is clear that the positivity constraint in \eqref{eq:dual_ground2} has exponential complexity $2^{O(d)}$ for direct implementation, although the number of optimization variables $\{X_{C_i}\}$ involved scales only polynomially in $d$. As mentioned in introduction, 
 it is actually QMA-hard to characterize those local Hamiltonians $\{H_{\gamma \d}\}$ in $\mc{L}_+$ because it is equivalent to the quantum marginal representability problem. 
 
 To elaborate, a set of local positive  functionals $\vr^{\sss V_{\gamma \d}} \in \mc{A}'_{\sss V_{\gamma \d}}$ for $\gamma < \d$ is jointly representable if there exists $\vr \in \mc{A}'_d$ such that $\vr \ge 0$ and $\vr|_{\mc{A}_{V_{\gamma \d}}} = \vr^{\sss V_{\gamma \d}}$. The representable marginals are always consistent, i.e.,
\begin{align} \label{eq:consistencygeneral}
    \vr^{V_{\gamma \d}}|_{\mc{A}_{V_{\gamma \d} \cap {V_{\gamma' \d'}}}} = \vr^{V_{\gamma' \d'}}|_{\mc{A}_{V_{\gamma \d} \cap {V_{\gamma' \d'}}}}\,,
\end{align}
for any $ \gamma < \d$ and $\gamma' < \d'$, where 
\begin{align*}
    {\mc{A}_{V_{\gamma \d} \cap {V_{\gamma' \d'}}}} := \big\l \{1\}\cup\{a_{i}\,;\,i\in C_{j},\ j \in V_{\gamma \d} \cap {V_{\gamma' \d'}}\} \big\r = \mc{A}_{V_{\gamma \d}} \bigcap \mc{A}_{V_{\gamma'\d'}}\,.
\end{align*}
By the bipolar theorem \cite{rockafellar1970convex}, 
 the dual cone of $\mc{L}_+$ in \eqref{eq:coneham} is given by the set of jointly representable marginals: 
\begin{align} \label{eq:joint_rep}
    \mc{L}^*_{+}
    &= \Big\{ \{\vr^{V_{\gamma\d}}\}_{\gamma < \d}  \,;\ \vr^{V_{\gamma\d}} \in \mc{A}'_{V_{\gamma \d}}\,,\ \sum_{\gamma < \d} \vr^{V_{\gamma \d}}(h_{\gamma \d}) \ge 0\,,\ \forall \{h_{\gamma \d}\}_{\gamma < \d} \in \mc{L}_+ \Big\} \notag \\ 
    &=\Big\{ \{\vr^{V_{\gamma\d}}\}_{\gamma < \d} \,;\ \{\vr^{V_{\gamma\d}}\}_{\gamma < \d} \ \text{are jointly representable}\Big\} \subset \prod_{\gamma < \d} \mc{A}'_{V_{\gamma \d}},
\end{align}
and vice versa. Analogous to \cite{liu2006consistency}, one can show that the joint representability \eqref{eq:joint_rep} is QMA hard, and so is \eqref{eq:dual_ground2}. 
As a sketch of the proof, this is because the ground-state energy problem \eqref{subtarget} can reduce to a convex optimization in variables $\{\vr^{V_{\gamma \d}}\} \in \mc{L}^*_{+}$,  which can be solved in polynomial time by the Bertsimas-Vempala algorithm or simulated annealing, if there is a membership oracle for the convex set $\mc{L}^*_{+}$. Then, the known QMA-hardness of the local Hamiltonian problem of fermions \cite{liu2006consistency,liu2007quantum} readily implies that the representability of $\{\vr^{V_{\gamma \d}}\}$ is QMA-hard. We remark that the above hardness result is only for the worst case, which leaves open the possibility of accurately computing the ground-state energy $E_0$ for a particular family of fermionic Hamiltonians in polynomial time.

We next adopt the SOS technique to relax the constraint set $\mc{L}_+$. The standard SOS hierarchy is built on the degree of polynomials involved in the SOS representation, which yields the famous RDM method when applied to the molecular ground state energy computation
\cite{mazziotti2002variational,mazziotti2004realization}; see also \eqref{eq:rdmhier} below. Given the sparsity of the Hamiltonian, we propose an alternative SOS hierarchy building on the dependent variables of polynomials, which, as we shall see in Section \ref{subsec:connection}, complements the RDM method and can be also interpreted as a quantum embedding theory.

We say that a Hamiltonian $h \in \mc{A}_d$ is a $k$th-order sum-of-squares (abbr.\,as $k$-SOS) with respect to clusters $\{V_\gamma\}_{\gamma \in [M]}$ if it can be written as 
\begin{align} \label{sos_rep}
    h = g_1^\dag g_1 + \cdots + g_m^\dag g_m\,,
\end{align}
where the monomials in each polynomial $g_i \in \mc{A}_d$ depend on at most $k$ groups $V_\gamma$, namely, each $g_i$ is spanned by the monomials of the form:
\begin{align} \label{eq:mono_basis}
    f_\alpha = a_{i_1}^\dag \cdots a_{i_s}^\dag a_{i_{s+1}}  \cdots a_{i_l}\,,
\end{align}
for some $s < l$ and 
\begin{align*}
    i_1 < \cdots < i_l \in  \bigcup\big\{C_j\,;\ j \ \text{is from at most}\ k \ \text{groups}\ V_\gamma \big\}\,.
\end{align*} 
Noting that any $k$-SOS $h$ is positive, and that 
$h \in \mc{A}_d$ is positive if and only if $h = g^\dag g$ for some $g \in \mc{A}_d$, it is straightforward to relax $ \mc{L}_{+}$ by the following convex cones: 
 \begin{align} \label{eq:defhk}
      \mc{L}^{(k)}_{+} := \Big\{\{h_{\gamma \d}\}_{\gamma < \d}   \,;\  \sum_{\gamma < \d}h_{\gamma \d} \ \text{is a}\ k\text{-SOS} \,,\ h_{\gamma \d}\in \mc{A}_{V_{\gamma \d}}  \Big\}, \q k \in [M]\,,
 \end{align} 
with
\begin{align*}
     \mc{L}^{(1)}_{+} \subset \dots \subset \mc{L}^{(M)}_{+} = \mc{L}_{+}\,.
\end{align*}
We thus obtain the relaxations for the problem \eqref{eq:dual_ground2}: 
\begin{align}  \label{eq:kthconst} 
    E_0^{(k)}[H] := \sup_{X_{C_i} \in \mc{A}^{{\rm sa}}_{C_i}} \Big\{\sum_i E_0\left[X_{C_i} + H_{C_i}\right] \,;\ \{H_{\gamma \d}\}_{\gamma < \d} \in  \mc{L}^{(k)}_{+}  \Big\}\,,
\end{align} 
with the hierarchy:
\begin{align*}
    E_0^{(1)} \le \cdots \le  E_0^{(M)} = E_0\,.
\end{align*}
We emphasize that $E^{(k)}_0$ does not depend on the splitting weights $w_{ij}^{\gamma \d}$ and $w_{i \sim j,i}^{\gamma \d}$ in \eqref{eq:split_hamil}, which would be clear from the formula \eqref{eq:embedding_hier} below.

\begin{remark}
To deal with more general Hamiltonians  
 \eqref{generalham}, it is necessary to consider the overlapping groups in \eqref{eq:group_v} such that similarly to \eqref{eq:partitionham}, 
\begin{equation*}
    H_{{\rm int}} - \sum_{i \in [d_c]} X_{C_i} = \sum_{\gamma < \d} H_{\gamma \d}\,.
\end{equation*}
Then, the SDP relaxations derived here and below can be extended straightforwardly. 
\end{remark}

\begin{remark} \label{rem1}
A higher-order relaxation $E^{\sss (k)}_0$ for the problem \eqref{subtarget} for a grouping of orbitals can always be viewed as a lower-order relaxation but with larger groups. For example, 
the $2$-SOS relaxation $E_0^{\sss (2)}$ with clusters $\{V_\gamma\}$ is nothing else than the $1$-SOS relaxation $E_0^{\sss (1)}$ with larger clusters 
$\{V_{\gamma \d}\}$. Thus, in principle, it suffices to consider the first-order approximations $E^{\sss (1)}_0$ with varying clusters to obtain a hierarchical relaxation for $E_0$. 
In Section \ref{sec:multiembedding}, we shall suggest a framework for optimally selecting the clusters to tighten the SDP relaxations  presented in this section.
\end{remark}

We now reformulate the relaxation $E_0^{(k)}$ as a
quantum embedding. 
From \eqref{eq:defhk} and \eqref{eq:kthconst}, we write 
\begin{align} \label{eq:kthconst_3}
     E_0^{(k)} = \sup_{\mycom{X_{C_i} \in \mc{A}^{{\rm sa}}_{C_i}}{\{h_{\gamma \d}\}_{\gamma < \d} \in \mc{L}_+^{(k)}}} \Big\{\sum_i E_0\left[X_{C_i} + H_{C_i}\right] \,;\   \sum_{\gamma < \d} H_{\gamma \d} =  \sum_{\gamma < \d} h_{\gamma \d}   \Big\}\,.
\end{align}
We introduce the Lagrange multiplier $l \in \mc{A}_d'$ for the constraint in \eqref{eq:kthconst_3} and find 
\small
\begin{align} \label{eq:relaxone}
      E_0^{(k)}[H] 
      & = \sup_{\mycom{X_{C_i} \in \mc{A}^{{\rm sa}}_{C_i}}{\{h_{\gamma \d}\}_{\gamma < \d} \in \mc{L}_+^{(k)}}} \inf_{l \in \mc{A}_d'} \bigg(\sum_i E_0\left[X_{C_i} + H_{C_i}\right] - \sum_{\gamma < \d} l^{\gamma \d} \big(H_{\gamma \d} - h_{\gamma \d}\big)\bigg) \notag \\ 
       & = \sup_{\mycom{X_{C_i} \in \mc{A}^{{\rm sa}}_{C_i}}{\{h_{\gamma \d}\}_{\gamma < \d} \in \mc{L}_+^{(k)}}}\inf_{\mycom{l^{\gamma \d} \in \mc{A}'_{V_{\gamma \d}}}{\{l^{\gamma \d}\}_{\gamma < \d}\,\text{are consistent}}} \bigg(\sum_i E_0\left[X_{C_i} + H_{C_i}\right] - \sum_{\gamma < \d} l^{\gamma \d} \big(H_{\gamma \d} - h_{\gamma \d}\big)\bigg)\,,
\end{align}
\normalsize
where
the linear functionals $l^{\gamma \d}$ in the first line denote the restrictions of $l \in \mc{A}_d'$ on $\mc{A}_{V_{\gamma \d}}$; the consistency of functionals $l^{\gamma \d}$ in the second line is a condition similar to \eqref{eq:consistencygeneral}: 
\begin{align} \label{eq:consis_const}
     l^{\gamma \d}|_{\mc{A}_{V_{\gamma \d} \cap {V_{\gamma' \d'}}}} = l^{\gamma' \d'}|_{\mc{A}_{V_{\gamma \d} \cap {V_{\gamma' \d'}}}}\,,\q \forall \gamma < \d\,, \gamma' < \d'\,.
\end{align}
It is worth emphasizing that the second equality in \eqref{eq:relaxone} is nontrivial. It was proved in \cite[Example 2.1]{maharam1972consistent} that the consistent linear maps on a family of linear subspaces are not necessarily jointly representable. However, in our setting,  the equivalence between the 
consistency and the representability of functionals $l^{\gamma \d} \in \mc{A}_{V_{\gamma \d}}'$ can be guaranteed by \cite[Theorem 5.1]{maharam1972consistent}.  


We consider the dictionary order on the pairs $\{(\gamma,\d)\}_{\gamma < \d}$ and let  $X^{\gamma \d, \gamma' \d'} \in \mc{A}_{V_{\gamma \d} \cap V_{\gamma'\d'}}$ be the multipliers for the 
constraints in \eqref{eq:consis_const} with $(\gamma, \d) < (\gamma', \d')$.  We proceed with \eqref{eq:relaxone} and derive
\small
\begin{align} \label{axueqq}
        E_0^{(k)}[H] 
        = \sup_{\mycom{X_{C_i} \in \mc{A}^{{\rm sa}}_{C_i},  
        \{h_{\gamma \d}\}_{\gamma < \d} \in \mc{L}_+^{(k)}
        }{X^{\gamma \d, \gamma' \d'} \in \mc{A}_{V_{\gamma \d} \cap V_{\gamma'\d'}}}}  \inf_{l^{\gamma \d} \in \mc{A}'_{V_{\gamma \d}}
        } \bigg (&\sum_i E_0\big[X_{C_i}  + H_{C_i}\big] - \sum_{\gamma < \d} l^{\gamma \d} \big(H_{\gamma \d} - h_{\gamma \d}\big)  \notag \\ &- \sum_{\mycom{\gamma < \d, \gamma' < \d'}{(\gamma,\d) < (\gamma',\d')}} \big(l^{\gamma \d} - l^{\gamma'\d'}\big)\big(X^{\gamma \d, \gamma' \d'}\big) \bigg)\,.
\end{align}
\normalsize
Defining $X_{\gamma' \d', \gamma \d} := - X_{\gamma \d, \gamma' \d'}$ for any $(\gamma, \d) < (\gamma', \d')$, then \eqref{axueqq} above gives
\small
\begin{align}  \label{eq:relaxation}
     E_0^{(k)}[H] 
      &  = \sup_{\mycom{X_{C_i} \in \mc{A}^{{\rm sa}}_{C_i}, \{h_{\gamma \d}\}_{\gamma < \d} \in \mc{L}_+^{(k)}}{X_{\gamma \d, \gamma' \d'} + X_{\gamma' \d'
        , \gamma \d} = 0}} \inf_{l^{\gamma \d} \in \mc{A}'_{V_{\gamma \d}}
        } \bigg (\sum_i E_0\big[X_{C_i}  + H_{C_i}\big] - \sum_{\gamma < \d}  l^{\gamma \d} \bigg(H_{\gamma \d} - h_{\gamma \d}  + \sum_{\mycom{\gamma' < \d'}{(\gamma,\d) \neq (\gamma',\d')}} X_{\gamma \d, \gamma' \d'} \bigg) \bigg) \notag \\
       & =  \sup_{\mycom{X_{C_i} \in \mc{A}^{{\rm sa}}_{C_i}, \{h_{\gamma \d}\}_{\gamma < \d} \in \mc{L}_+^{(k)}}{X_{\gamma \d, \gamma' \d'} + X_{\gamma' \d'
        , \gamma \d} = 0}} \bigg \{\sum_i E_0\big[X_{C_i}  + H_{C_i}\big]\,;\,  H_{\gamma \d}- h_{\gamma \d}  + \sum_{\mycom{\gamma' < \d'}{(\gamma,\d) \neq (\gamma',\d')}} X_{\gamma \d, \gamma' \d'} = 0\ \ \text{on}\ \mc{A}_{V_{\gamma \d}}\bigg\}\,.
\end{align}
\normalsize
We can let both $h_{\gamma \d}$ and $X_{\gamma \d, \gamma' \d'}$ above be Hermitian to reduce free variables without changing the optimal value. 
To interpret the SDP relaxation \eqref{eq:relaxation} as a quantum embedding method for computing $E_0$, it suffices to note that it is an optimization of a sum of local ground-state energy problems on 
fragments $C_i$, where the on-site Hamiltonians $X_{C_i}$ satisfy the local constraints on $V_{\gamma \d}$ with \emph{communication variables} $X_{\gamma \d, \gamma' \d'}$ that connect different clusters $\{V_{\gamma \d}\}_{\gamma < \d}$, and these local constraints are glued together by a global one $\{h_{\gamma \d}\}_{\gamma < \d} \in \mc{L}_+^{\sss (k)}$. In Remark \ref{rem:inter_eff} below, an alternative and clearer quantum embedding interpretation will be provided from the perspective of local effective Hamiltonians.

\subsection{Efficient embedding schemes} \label{subsec:embedding}
In this subsection, we shall derive some practical variational embedding schemes based on the general hierarchy \eqref{eq:relaxation}. 
For ease of exposition, we denote by $\mc{W} = \{f_1, \ldots, f_{|\mc{W}|}\}$ an ordered list of polynomials in $\mc{A}_d$ and 
define the associated matrix $\ff^{\sss \mc{W}}$ by 
\begin{align} \label{def:matrix_list}
\ff^{\mc{W}}_{ij} := f^\dag_i f_j\,,\q  \ 1 \le  i,j \le |\mc{W}|\,.
\end{align}
We introduce the convex cone in $\prod_{\gamma < \d} \mc{A}_{V_{\gamma \d}}$ corresponding to $\mc{W}$:
\begin{align} \label{eq:deflw}
      \mc{L}_{+}(\mc{W}) := \Big\{\{h_{\gamma \d}\}_{\gamma < \d}   \,;\  \sum_{\gamma < \d}h_{\gamma \d} =  \tr (Y\ff^{\mc{W}})\ \text{for}\ Y \ge 0\ \text{of size}\ |\mc{W}| \t |\mc{W}| \Big\}\,.
\end{align}
By definitions \eqref{eq:defhk} and \eqref{eq:deflw}, we have $\mc{L}_+^{(k)} = \mc{L}_+(\mc{W}_k)$ with $\mc{W}_k$ given by the spanning monomials in \eqref{eq:mono_basis}. Then, 
it follows from \eqref{eq:kthconst} that 
\small
\begin{align}  \label{eq:kthconst_2} 
    E_0^{(k)} = \sup_{X_{C_i} \in \mc{A}^{{\rm sa}}_{C_i}} \Big\{\sum_i E_0\left[X_{C_i} + H_{C_i}\right] \,;\   \sum_{\gamma < \d} H_{\gamma \d} =  \tr (Y\ff^{\mc{W}_k})\,\ \text{for}\  Y \ge 0 \Big\}\,.
\end{align} 
\normalsize
To efficiently compute $E_0^{\sss (k)}$ (instead of solving \eqref{eq:kthconst_2} with black-box methods), it is necessary to characterize the admissible set $\mc{L}_+^{\sss (k)}$. 
Note that, given a partition of orbitals \eqref{eq:group_v}, a $k$-SOS is a $2k$-cluster Hamiltonian (that is, we can write it as $\sum_i H_i$ with each $H_i$ depending on at most $2k$ clusters), while the Hamiltonian $H_{\gamma \d}$ is a $2$-cluster one. The $k$-SOS ($k \ge 2$) constraint in \eqref{eq:kthconst} means that we must carefully choose $g_j$ in \eqref{sos_rep} such that the $2i$-cluster terms with $i > 1$ in $\sum_{j} g_j^{  \dag} g_j$ cancel with each other. Unfortunately, it is 
generally very difficult to find all such cancellations, and obtain the full characterization of $\mc{L}_+^{\sss (k)}$, except when $k = 1$. 

By abuse of notations, let $\{f_{\gamma,\alpha}\}_{\alpha \in [4^{|V_\gamma|}]}$ be the basis for the algebra $\mc{A}_{V_\gamma}$, and define 
\begin{equation*}
 \mc{W}_1 = \{f_{1,1},\ldots,f_{1,4^{|V_1|}},\ldots, f_{M,1},\ldots,f_{M,4^{|V_M|}}\}\,.   
\end{equation*}
Then, $\ff^{\mc{W}_1}$ and $Y$ in \eqref{eq:deflw} are block matrices
of size $\big(\sum_\gamma 4^{|V_\gamma|}\big)\times \big(\sum_\gamma 4^{|V_\gamma|}\big)$ with blocks $\ff^{\mc{W}_1}_{\gamma \d}$ for $\gamma, \d \in [M]$ given by 
\begin{align*}
    (\ff^{\mc{W}_1}_{\gamma \d})_{\alpha \beta} = f_{\gamma,\alpha}^\dag f_{\d,\beta}\,, \q  1 \le \alpha \le 4^{|V_\gamma|}\,, \ 1 \le \beta  \le 4^{|V_\d|}\,.
\end{align*}
We hence have, for any $Y \ge 0$, 
\begin{align} \label{eq:blocksos}
    \tr(Y \ff^{\mc{W}_1}) = \sum_{\gamma, \d} \sum_{\alpha,\beta} (Y_{\d \gamma})_{\beta \alpha} f_{\gamma,\alpha}^\dag f_{\d,\beta} = \sum_{\gamma < \d} \w{H}_{\gamma \d}\,,
\end{align}
with 
\begin{align} \label{eq:blocksos_2}
    \w{H}_{\gamma \d} := \sum_{\alpha = 1}^{4^{|V_\gamma|}} \sum_{\beta = 1}^{4^{|V_\d|}} & (Y_{\d \gamma})_{\beta \alpha} f_{\gamma,\alpha}^\dag f_{\d,\beta} + \big((Y_{\d \gamma})_{\beta \alpha} f_{\gamma,\alpha}^\dag f_{\d,\beta}\big)^{\dag} \notag \\
     & + \sum_{\alpha,\beta = 1}^{4^{|V_\gamma|}} \ww_{\gamma\d,\gamma} (Y_{\gamma \gamma})_{\beta \alpha} f_{\gamma,\alpha}^\dag f_{\gamma,\beta} + \sum_{\alpha,\beta = 1}^{4^{|V_\d|}} \ww_{\gamma\d,\d} (Y_{\d \d})_{\beta \alpha} f_{\d,\alpha}^\dag f_{\d,\beta}\,,
\end{align}
where the weights satisfy $\sum_{\gamma < \d} 
\ww_{\gamma\d,\gamma} + \sum_{\d < \gamma} \ww_{\d\gamma,\gamma} = 1$ for any $\gamma$. By \eqref{eq:relaxation}, we have
\small
\begin{align} \label{scheme1}
     E_0^{(1)}[H] 
      =  \sup_{\mycom{X_{C_i}, Y \ge 0}{X_{\gamma \d, \gamma' \d'} + X_{\gamma' \d'
        , \gamma \d} = 0}} \bigg \{\sum_i E_0\big[X_{C_i}  + H_{C_i}\big]\,;\  H_{\gamma \d}- \w{H}_{\gamma \d}  + \sum_{\mycom{\gamma' < \d'}{(\gamma,\d) \neq (\gamma',\d')}} X_{\gamma \d, \gamma' \d'} = 0 \bigg\}\,,
\end{align}
\normalsize
where all the involved optimization variables are self-adjoint, and $H_{\gamma \d}$ and $\w{H}_{\gamma \d}$ are given in \eqref{eq:split_hamil} and \eqref{eq:blocksos_2}, depending on $\{X_{C_i}\}$ and $Y$,  respectively.

It is helpful to realize that the definition \eqref{eq:kthconst} of $E^{(k)}_0$ depends on 
$\mc{L}_+^{(k)}$, which is fully characterized by the polynomial list $\mc{W}_k$. In practice, as we can see in schemes \eqref{eq:dualscheme2} and \eqref{eq:t1embedding} below, instead of considering the whole list $\mc{W}_k$, a suitable choice of subsets of $\mc{W}_k$ suffices to yield accurate approximations of $E_0[H]$ with an efficient implementation. 
We consider the case $k = 2$. We introduce the lists $\mc{W}_{2,\gamma \d} = \{f_{\gamma\d, 1},\ldots, f_{\gamma\d, 4^{|V_{\gamma \d}|} }\} $ with $f_{\gamma \d, \alpha}$ being the basis of $\mc{A}_{V_{\gamma \d}}$, and define 
\small
\begin{align} \label{eq:defl21}
     \mc{L}_+^{(1,2)}   := & \Big\{\{h_{\gamma \d}\}_{\gamma < \d}   \,;\  \sum_{\gamma < \d}h_{\gamma \d} =  \tr (Y\ff^{\mc{W}_1}) + \sum_{\gamma < \d} \tr (Y_{\gamma \d} \ff^{\mc{W}_{\gamma \d}})\,,\ Y, Y_{\gamma \d} \ge 0 \Big\} \notag \\
    = & \Big\{\{h_{\gamma \d}\}_{\gamma < \d}   \,;\  \sum_{\gamma < \d}h_{\gamma \d} =  \tr (Y\ff^{\mc{W}_1}) + \sum_{\gamma < \d} g_{\gamma \d}\,,\ Y \ge 0,\ 0\le g_{\gamma \d } \in\mc{A}^{{\rm sa}}_{V_{\gamma \d}}   \Big\}\,,
\end{align}
\normalsize
where $Y$ and $Y_{\gamma \d}$ are of size $\big(\sum_\gamma 4^{|V_\gamma|}\big)\times \big(\sum_\gamma 4^{|V_\gamma|}\big)$ and $4^{|V_{\gamma \d}|} \times 4^{|V_{\gamma \d}|}$, respectively. 
With the help of $\mc{L}_+^{\sss (1,2)}$, we have the following relaxation, which is tighter than \eqref{scheme1}: 
\small
\begin{align} \label{scheme2}
         E_0^{(1,2)}[H] 
        :=  & \sup_{\mycom{X_{C_i} \in \mc{A}^{{\rm sa}}_{C_i}, \{h_{\gamma \d}\}_{\gamma < \d} \in \mc{L}_+^{(1,2)}}{X_{\gamma \d, \gamma' \d'} + X_{\gamma' \d'
        , \gamma \d} = 0}} \bigg \{\sum_i E_0\big[X_{C_i}  + H_{C_i}\big]\,;\  H_{\gamma \d}- h_{\gamma \d}  + \sum_{\mycom{\gamma' < \d'}{(\gamma,\d) \neq (\gamma',\d')}} X_{\gamma \d, \gamma' \d'} = 0 \bigg\} \notag \\
        = & \sup_{\mycom{X_{C_i}, Y \ge 0}{X_{\gamma \d, \gamma' \d'} + X_{\gamma' \d'
        , \gamma \d} = 0}} \bigg \{\sum_i E_0\big[X_{C_i}  + H_{C_i}\big]\,;\  H_{\gamma \d}- \w{H}_{\gamma \d}  + \sum_{\mycom{\gamma' < \d'}{(\gamma,\d) \neq (\gamma',\d')}} X_{\gamma \d, \gamma' \d'} \ge 0 \bigg\}.
\end{align}
\normalsize
Again, these relaxations are independent of the splitting weights in the local Hamiltonians $H_{\gamma \d}$ and $\w{H}_{\gamma \d}$ in \eqref{eq:split_hamil} and \eqref{eq:blocksos_2}, due to the existence of communication variables $X_{\sss \gamma \d, \gamma' \d'}$.

\begin{figure}[!htbp]
  \centering
  \includegraphics[width=0.4\textwidth]{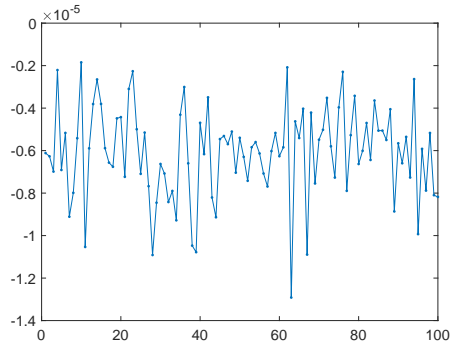}
  \caption{Errors $(E_0^{(1,2)'}-E_0^{(1,2)})/N$ for the toy Hamiltonian considered in Section \ref{sec:adaptive_embedding} (i.e., TFI model \eqref{def:tfimodel} with the underlying graph in Figure \ref{fig:graph1} and coefficients $h_i = J_{ij} = 1$) with the clusters $C_i = \{i\}$ in \eqref{eq:index_cluster_fermi} and $V_\gamma = \{\gamma\}$ in \eqref{eq:group_v}
  and 100 instances of randomly generated splitting weights in \eqref{eq:split_hamil} and \eqref{eq:blocksos_2}.} 
  \label{fig:weight} 
\end{figure}

One can omit the variables $X_{\sss \gamma \d, \gamma' \d'}$ and obtain a less tight relaxation:
\small
\begin{align} \label{scheme3}
         E_0^{(1,2)'}[H] 
        = & \sup_{X_{C_i}, Y \ge 0} \bigg \{\sum_i E_0\big[X_{C_i}  + H_{C_i}\big]\,;\  H_{\gamma \d}- \w{H}_{\gamma \d}  \ge 0 \bigg\}\,,
\end{align}
\normalsize
which has a simpler implementation and reduced computational cost. By definition, this scheme would depend on the weights in $H_{\gamma \d}$ and $\w{H}_{\gamma \d}$. However, we numerically observe that it is very robust with respect to the choice of weights and can produce nearly the same result as $E_0^{\sss (1,2)}$ in \eqref{scheme2}, at least for our tested Hamiltonians; see Figure \ref{fig:weight} below for numerical evidence. 

For a larger $k$, it is non-trivial to select subsets $\mc{W} $ from $\mc{W}_k$ such that the associated $\mc{L}_+(\mc{W})$ can be easily characterized. In the next section, we will exploit the relations with the variational RDM method, which opens the possibility of utilizing the well-studied 2-RDM representability conditions to tighten the embedding schemes; see \eqref{eq:t1embedding} and \eqref{eq:rela_rdm}.

\begin{remark}
Since any $\mc{W}_k$ for $k \ge 2$ includes the subsets $\mc{W}_{2,\gamma \d}$, by the same derivation as that of \eqref{scheme2}, we can replace the equality in the constraint in $E_0^{\sss (k)}$ \eqref{eq:relaxation} by the inequality, i.e., 
\small
\begin{align} \label{scheme_general}
      E_0^{(k)}[H] 
       & =  \sup_{\mycom{X_{C_i} \in \mc{A}^{{\rm sa}}_{C_i}, \{h_{\gamma \d}\}_{\gamma < \d} \in \mc{L}_+^{(k)}}{X_{\gamma \d, \gamma' \d'} + X_{\gamma' \d'
        , \gamma \d} = 0}} \bigg \{\sum_i E_0\big[X_{C_i}  + H_{C_i}\big]\,;\  H_{\gamma \d}- h_{\gamma \d}  + \sum_{\mycom{\gamma' < \d'}{(\gamma,\d) \neq (\gamma',\d')}} X_{\gamma \d, \gamma' \d'} \ge 0 \bigg\}\,,
\end{align}
\normalsize
which will be useful later. 
\end{remark}

\begin{remark}[(\emph{Sublevel embedding})] \label{rem:sublevel}
Note that
the SDP relaxations $E_0^{\sss (k)}[H]$ in \eqref{scheme_general} (also those in \eqref{scheme2} and \eqref{scheme3}) scale exponentially in cluster size $|V_\gamma| := \sum_{\sss  j \in V_\gamma} |C_j|$. When $|V_\gamma|$ becomes large, the positivity constraints on $\{V_{\gamma \d}\}$ make the computation of $E_0^{\sss (k)}[H]$ intractable. A straightforward way to remedy this issue is to introduce a sublevel SOS hierarchy to further relax the local positivity conditions. The classical sublevel Lasserre’s hierarchy has been recently explored in \cite{josz2018lasserre,chen2022sublevel}. Clearly, a sublevel relaxation would depend on how we divide $V_{\gamma \d}$ into smaller clusters and the choice of the monomials involved in the SOS representation, as well as the number of sublevels. Such flexibility helps to balance the accuracy and the computational complexity, while it also raises the question of how to choose the sublevel clusters, monomials, and depth optimally. We leave detailed investigations on the sublevel variational quantum embedding for future work. 
\end{remark}

\subsection{Connections with 
existing methods} \label{subsec:connection}
We will relate our SOS-based quantum embedding method with some existing ones for lower bounds: the variational embedding by Lin and Lindsey \cite{lin2022variational}, the RDM method \cite{mazziotti2007reduced}, and the Anderson bounds \cite{anderson1951limits,wittmann1993bounds}.

As a preparation, we write the dual problem of \eqref{subtarget} as follows:
\begin{align} \label{eq:fulldual}
    E_0[H] = \sup_{\lad \in \R} \{\lad\,;\ H - \lad \ge 0\}\,.
\end{align}
Then we recall 
the $k$th-order relaxation $E_0^{\sss (k)}$ in \eqref{eq:kthconst} and find, by \eqref{eq:fulldual}, 
\begin{align} \label{eq:kthconst_4}
      E_0^{(k)}[H] = \sup_{X_{C_i}  \in \mc{A}^{{\rm sa}}_{C_i},\, \lad_i \in \R} \Big\{\sum_i \lad_i \,;
       \sum_{1 \le i < j \le d_c} H_{C_{ij}} & - \sum_{i \in [d_c]} X_{C_i}  \  \text{is a}\ k\text{-SOS}\,, \notag \\ & X_{C_i} + H_{C_i} - \lad_i \ge 0 \ \text{for all}\ i 
       \Big\}\,.
\end{align}
We claim that there holds 
\begin{align} \label{eq:embedding_hier}
     E_0^{(k)}[H] = \sup_{\lad \in \R} \Big\{\lad \,;
       H - \lad \ \text{is a}\ k\text{-SOS}
       \Big\}\,.
\end{align}
Indeed, the direction ($\ge$) simply follows from  taking $X_{C_i} = \lad_i - H_{C_i}$ in \eqref{eq:kthconst_4} and the formula \eqref{hamiltonian_decom}. For the other direction ($\le$), it suffices to note that a positive polynomial $X_{C_i} + H_{C_i} - \lad_i$ is always a $k$-SOS, so that the constraint in \eqref{eq:kthconst_4} implies that $H - \sum_{i} \lad_i$ is a $k$-SOS. Therefore, the claim holds. Similarly, recalling \eqref{eq:defl21} and \eqref{scheme2},  
the relaxation $E^{\sss (1,2)}_0$ can be reformulated as
\begin{align} \label{eq:dualscheme2}
        E_0^{(1,2)}[H] 
     & = \sup_{\mycom{X_{C_i} \in \mc{A}^{{\rm sa}}_{C_i}}{0\le g_{\gamma \d } \in\mc{A}^{{\rm sa}}_{V_{\gamma \d}}  }} \bigg \{\sum_i E_0\big[X_{C_i}  + H_{C_i}\big]\,;\  
       \sum_{i < j} H_{C_{ij}} - \sum_{i } X_{C_i} -\sum_{\gamma < \d} g_{\gamma \d}\ \text{is a}\ 1\text{-SOS}
        \bigg\} \notag \\
    & = \sup_{0\le g_{\gamma \d } \in\mc{A}^{{\rm sa}}_{V_{\gamma \d}}, \lad \in \R} \bigg\{\lad \,;\ 
    H - \lad  = 1\text{-SOS} + \sum_{\gamma < \d} g_{\gamma \d}\bigg\}\,.
\end{align}

\begin{remark} \label{rem3}
A simple consequence of formulations \eqref{eq:fulldual} and \eqref{eq:embedding_hier} is that $E^{\sss (k)}_0$ is exact (not a lower bound of $E_0$) if and only if for some $a \in \R$, $H - a$ is a $k$-SOS. 
\end{remark}

\begin{remark} \label{rem4}   
One can derive the SDP relaxation hierarchy starting from \eqref{eq:embedding_hier} alternatively, instead of from \eqref{eq:kthconst}.
We choose to include the formulations relating to the multi-marginal quantum OT mainly for completeness and the potential discussion on entropic regularization \cite{lindsey2023fast,feliciangeli2023non}. 
\end{remark}

We next show the equivalence between the relaxation $E^{\sss (1,2)}_0$ in \eqref{scheme2} and the one in \cite{lin2022variational}. For simplicity, we  consider the non-overlapping groups $V_\gamma = \{\gamma\}$ in \eqref{eq:group_v} (also note that the case of overlapping clusters was only sketched in \cite{lin2022variational}). We recall the dual formulation of the two-marginal SDP in \cite[Section 5.4]{lin2022variational} with respect to the clusters $\{C_j\}$ in \eqref{eq:index_cluster_fermi}:
\small
\begin{eqnarray}  \label{embddinglin}
E'_0[H] = \underset{Y \ge 0,\,\lambda_i \in \R,\, A_{ij} \in \mc{A}_{C_{i}}^{{\rm sa}},\, B_{ij}\in \mc{A}_{C_{j}}^{{\rm sa}}}{\mbox{sup}} \quad &  & \sum_{i} \lad_i\\
\mbox{subject to} \quad &  & H_{C_i} - h_{i}(Y_{ii})-\lambda_{i}+\sum_{j>i}A_{ij}+\sum_{j<i}B_{ji}=0,\quad i=1,\ldots, d_c, \notag \\
 &  & A_{ij} +  B_{ij} \le H_{C_{ij}} - h_{ij}(Y_{ij}),\quad1\leq i<j\leq d_c\,, \notag 
\end{eqnarray}
\normalsize
where $Y$ is a block matrix with blocks $Y_{ij} \in \C^{4^{|C_i|} \t 4^{|C_j|}}$ for $i,j \in [d_c]$, and the maps $h_i(\dd): \C^{4^{|C_i|} \t 4^{|C_i|}} \to \mc{A}_{C_i}$ and $h_{ij}(\dd): \C^{4^{|C_i|} \t 4^{|C_j|}} \to \mc{A}_{C_{ij}}$ are given by
\begin{align*}
h_{i}(X)=\sum_{\alpha,\beta}\overline{X}_{\alpha\beta}\,f_{i,\alpha}^{\dagger}f_{i,\beta}\,,\quad h_{ij}(X)=\sum_{\alpha,\beta}\overline{X}_{\alpha\beta}\,f_{i,\alpha}^{\dagger} f_{j,\beta}+ \big(\overline{X}_{\alpha\beta}\,f_{i,\alpha}^{\dagger} f_{j,\beta}\big)^\dag\,.
\end{align*}
Here, $\{f_{i,\alpha}\}$ is the basis of $\mc{A}_{C_i}$. To show $E_0'[H]  = E_0^{\sss (1,2)}[H]$, we define 
\begin{align} \label{auxeq2}
    g_{ij} := H_{C_{ij}} - h_{ij}(Y_{ij}) - (A_{ij} +  B_{ij}) \ge 0\,,
\end{align}
and note that there hold $\mc{A}_{V_{\gamma \d}} = \mc{A}_{C_{ij}}$ by $V_\gamma = \{\gamma\}$, and  
\begin{align} \label{auxeq4}
    \sum_{1 \le i < j \le d_c} h_{ij}(Y_{ij}) + \sum_{i \in [d_c]} h_i(Y_{ii}) = \tr(Y \ff^{\mc{W}_1})\,,
\end{align}
where $\tr(Y \ff^{\mc{W}_1})$ is given in \eqref{eq:blocksos}. It then follows from \eqref{auxeq2} that the two constraints in \eqref{embddinglin} are equivalent to the following one: for any $i \in [d_c]$, 
\begin{align} \label{auxeq3}
  \sum_{i < j} B_{ij} - \sum_{j<i}B_{ji} =   H_{C_i} + \sum_{i < j} H_{C_{ij}} - h_{i}(Y_{ii}) -\lambda_{i} - \sum_{i < j} \big(g_{ij} + h_{ij}(Y_{ij})\big)\,,
\end{align}
which can be viewed as a linear equation in $\{B_{ij}\}_{i<j}$. By basic algebra, we obtain that the equation \eqref{auxeq3} is consistent (equivalently, the constraint is feasible) if and only if 
\begin{align*}
    \sum_{i} \bigg( H_{C_i} + \sum_{i < j} H_{C_{ij}} - h_{i}(Y_{ii}) -\lambda_{i} - \sum_{i < j} \big(g_{ij} + h_{ij}(Y_{ij})\big)\bigg) = 0\,.
\end{align*}
This, along with \eqref{eq:dualscheme2} and \eqref{auxeq4}, allows us to conclude that 
\begin{align*}
    E_0'[H] = \sup_{0\le g_{ij} \in\mc{A}^{{\rm sa}}_{C_{ij}}, \lad_i \in \R} \bigg\{\sum_i\lad_i \,;
    H -\sum_{i < j} g_{ij} - \sum_i \lad_i \ \text{is a}\ 1\text{-SOS}\bigg\} = E_0^{(1,2)}[H]\,.
\end{align*}

\medskip 

We now discuss the connections with the RDM method and some interesting implications. We first briefly review the RDM method for the model \eqref{eq:hamiltonian_general} for ease of exposition. We say that $h \in \mc{A}_d$ is a degree-$k$ sum of squares (abbr.\,as degree $k$-SOS) if it can be written as $h = g_1^\dag g_1 + \cdots + g_m^\dag g_m$ for some polynomials $g_i \in \mc{A}_d$ of degree at most $k$. The hierarchy for the RDM method is then given by 
\begin{align} \label{eq:rdmhier}
E_0^{(k),{\rm RDM}}[H] = \sup_{\lad \in \R} \Big\{\lad \,;
       H - \lad \ \text{is a degree}\ k\text{-SOS}
       \Big\}\,.
\end{align}
To be precise, we define the 1-RDM and 2-RDM elements for a quantum state $\vr \in \mc{D}(\mc{A}_d)$ by 
\begin{align*}
\gamma_{ij} := \vr(a_i^\dag a_j)\,,\q  \Gamma_{ijkl} := \vr(a_i^\dag a_j^\dag a_l a_k)\,,
\end{align*}
which satisfy the following \emph{necessary consistency conditions}, from CARs \eqref{eq:commu},
\begin{align} \label{cons1}
    \overline{\gamma_{ij}} = \gamma_{ji}\,,\q \overline{\Gamma_{ijkl}} = \Gamma_{klij}\,,\q \sum_{i} \gamma_{ii} \le d\,,\q \sum_{ij} \Gamma_{ijij} \le d (d-1)\,,
\end{align}
and 
\begin{align} \label{cons2}
   \Gamma_{ijkl} = - \Gamma_{jikl} = -\Gamma_{ijlk} = \Gamma_{jilk}\,.
\end{align}
By the physical assumptions of the superselection rules and the particle-number conservation for the fermionic Hamiltonian, the expectation value $\vr(p)$ of a monomial $p$ vanishes unless $p$ is even and involves the same number of $a_i^\dag$ and $a_i$; see \cite{baumgratz2012lower} for more details. Then it is easy to show that the dual problem of $E_0^{\sss (2),{\rm RDM}}$ \eqref{eq:rdmhier} gives the 2-RDM method with the so-called $P,Q,G$ conditions: 
\begin{eqnarray}  \label{2rdmdqg}
E_0^{(2),{\rm RDM}}[H] = \underset{\gamma,\Gamma}{\mbox{inf}} \quad &  & \sum_{ij} t_{ij} \gamma_{ij} + \sum_{ijkl}v_{ijkl} \Gamma_{ijkl}
\\
\mbox{subject to} \quad &  & \text{consistency conditions}\ \eqref{cons1}\ \text{and}\ \eqref{cons2}\,, \notag \\
 &  & \gamma \ge 0\,,\q I - \overline{\gamma} \ge 0\,, \notag \\ 
 &  & \Gamma \ge 0\,,\q Q \ge 0\,,\q G \ge 0\,. \notag
\end{eqnarray}
Here, the matrices $Q$ and $G$ are defined by 
\begin{align}\label{def:matrix}
    Q_{ijkl} = \vr(a_{i} a_{j} a_{l}^\dag a_{k}^\dag)\,,\q 
G_{ijkl} = \vr(a_{i}^\dag a_{j} a_{l}^\dag a_{k}) - \vr(a_i^\dag a_j) \vr(a_l^\dag a_k)\,,
\end{align}
which, by CARs \eqref{eq:commu}, can be represented by $1$-RDM and $2$-RDM elements $\gamma_{ij}$ and $\Gamma_{ijkl}$. We should remark that the formulation of $G$ in \eqref{def:matrix} dates back to the early work of 
\cite{garrod1964reduction} by Garrod 
and Percus in 1964, while in recent RDM literature, the matrix $G$ is usually given by $G_{ijkl} 
= \vr(a_{i}^\dag a_{j} a_{l}^\dag a_{k})$, which has the same eigenvalues as those of $G$ in \eqref{def:matrix} for the states with fixed particle number \cite{erdahl1979two}. We can introduce some fragments of the higher-order degree SOS to obtain tighter approximations. A simple but very effective choice is
\begin{align} \label{eq:poly1}
P_{T_{1,1}} = \sum_{jkl} b_{jkl} a_j^\dag a_k^\dag a_l^\dag\,,\q P_{T_{1,2}} = \sum_{jkl} \overline{b_{jkl}} a_j a_k a_l\,,
\end{align}
and 
\begin{align} \label{eq:poly2}
    P_{T_{2,1}} = \sum_{jkl} c_{jkl} a_j^\dag a_k^\dag a_l\,,\q P_{T_{2,2}} = \sum_{jkl} \overline{c_{jkl}} a_j a_k a_l^\dag\,,
\end{align}
with $b_{jkl}, c_{jkl} \in \C$. The associated relaxation is then defined by 
\begin{align} \label{2rdmt1t2}
    E_0^{(2,3),{\rm RDM}}[H] = \sup_{\lad \in \R} \Big\{\lad \,;\ 
       H - \lad = &\ \text{a degree}\ 2\text{-SOS} + \sum_{i,j = 1}^2 P_{T_{i,j}}^\dag P_{T_{i,j}} \notag \\
       & \text{with}\ \{P_{T_{i,j}}\}\ \text{given in}\ \eqref{eq:poly1}\ \text{and}\ \eqref{eq:poly2} 
       \Big\}\,,
\end{align}
which is tighter than $E_0^{(2),{\rm RDM}}$. Note from  \eqref{eq:commu} that the polynomial $\sum_{i,j = 1}^2 P_{T_{i,j}}^\dag P_{T_{i,j}}$ in \eqref{2rdmt1t2} is of degree at most $4$. One can prove that the dual formulation of \eqref{2rdmt1t2} is the problem \eqref{2rdmdqg} with additional $T_1$ and $T_2$ conditions on the RDMs $\gamma$ and $\Gamma$ \cite{zhao2004reduced,fukuda2007large,mazziotti2005variational,mazziotti2006variational}. We refer the interested readers to \cite{mazziotti2012structure,mazziotti2012significant} for a systematic construction of the hierarchical $2$-RDM relaxations for the ground-state energy problem, which generally takes the following form: for $k \ge 3$, 
\begin{align} \label{rdmgeneral}
 E_0^{(2,k),{\rm RDM}}[H] = \sup_{\lad \in \R} \Big\{\lad \,;\ 
       H - \lad = \text{a degree}\ 2\text{-SOS} + \sum_{j} P_{{k,j}}^\dag P_{{k,j}} \Big\}\,,
\end{align}
with $P_{k,j}$ being the linear combination of some fragments of degree-$k$ monomials such that $\sum_{j} P_{{k,j}}^\dag P_{{k,j}}$ is of degree at most $4$. 

Note from \eqref{eq:embedding_hier} and \eqref{eq:rdmhier} that the hierarchies of the  variational embedding and the RDM method are in principle different: one is based on the support (locality) of the SOS representation while the other one is based on the degree of SOS. Nevertheless, in specific settings, they are comparable with each other. 
For the sake of clarity, we focus on the model \eqref{eq:hamiltonian_general} and let $C_i = \{i\}$ in \eqref{eq:index_cluster_fermi} and $V_\gamma = \{i,j\}$ for $i < j$ in \eqref{eq:group_v}. Recall that $\mc{A}_{C_i} = {\rm span}\{1,a_i,a_i^\dag,a_i^\dag a_i\}$. By the formulation \eqref{eq:embedding_hier} of $E_0^{\sss (k)}$ with $k = 1$, we obtain
\small
\begin{eqnarray}  \label{embedding2}
  E_0^{(1)}[H] = \underset{\lad \in \R}{\mbox{sup}} \quad &  & \lad 
\\
\mbox{subject to} \quad &  & H - \lad = g^\dag g \ \text{with}\ g \ \text{of the form}: \notag \\
 &  &  g = \d + \sum_i (\alpha_i a_i  + \beta_i a_i^\dag) + \sum_{ij} (\eta_{ij} a_i a_j + \tau_{ij} a_i^\dag a_j + \ep_{ij} a_i^\dag a_j^\dag) \notag \\ 
 &  & \qq + \sum_{ij} (b_{ij} a_i^\dag a_j^\dag a_j +  c_{ij} a_i a_j^\dag a_j + f_{ij} \h{n}_i \h{n}_j) \notag \\ 
 & & \qq  \text{where}\ \d, \alpha_i,  \beta_i, \eta_{ij}, \tau_{ij}, \ep_{ij}, b_{ij}, c_{ij}, f_{ij} \in \C\,. \notag 
\end{eqnarray}
\normalsize
Then, several interesting consequences are in order. First, since the admissible $g$ in \eqref{embedding2} include all the degree-$2$ polynomials in $\mc{A}_d$, the relaxation $E_0^{\sss (1)}$ (and also $E_0^{\sss (1,2)}$) is a tighter lower bound than $E_0^{\sss (2),{\rm RDM}}$ in \eqref{2rdmdqg}: 
\begin{align} \label{eq:compare}
    E_0^{(2),{\rm RDM}}[H] \le  E_0^{(1)}[H] \le  E_0^{(1,2)}[H] \le E_0[H]\,.
\end{align} 

Second, for the non-interacting case $H = \sum_{ij} t_{ij} a_i^\dag a_j$ (i.e., the free fermion), it holds that, for any clusters $\{C_i\}$ and groupings $\{V_\gamma\}$,
\begin{align*}
E_0^{(1)}[H] \ge  E_0^{1,{\rm RDM}}[H] = E_0[H]\,,
\end{align*}
that is, $E_0^{\sss (1)}[H]$ (and hence all of the higher-order relaxations) is exact. Alternatively, this fact can follow from Remark \ref{rem3}, by diagonalizing the matrix $T = \{t_{ij}\}$ and finding that $H - \sum_{\ep_j < 0} \ep_j$ is a $1$-SOS, where $\ep_j$ are eigenvalues of $T$.

Third, since any degree-$4$ Hamiltonian $H$ can be written as the sum of local Hamiltonians $\sum_{\gamma < \d} h_{\gamma \d}$ with $h_{\gamma \d} \in \mc{A}_{V_{\gamma \d}}^{{\rm sa}}$ for $V_\gamma = \{i,j\}$, theoretically, one can combine \emph{any} RDM scheme (say, $E_0^{\sss (2,k),{\rm RDM}}$ in \eqref{rdmgeneral}) into the variational embedding to tighten the relaxation. For example, we consider the $T_1$ and $T_2$ conditions in the RDM method. We introduced the following lists associated with \eqref{eq:poly1} and \eqref{eq:poly2} in the dictionary order: 
\small
\begin{align*}
\mc{W}_{T_{1,1}} = \{a_1^\dag a_1^\dag a_1^\dag,  a_1^\dag a_1^\dag a_2^\dag, \ldots, a_d^\dag a_d^\dag a_d^\dag\}\,,\q \mc{W}_{T_{1,2}} = \{a_1 a_1 a_1,  a_1 a_1 a_2, \ldots, a_d a_d a_d\}\,,
\end{align*}
\normalsize
and 
\small
\begin{align*}
\mc{W}_{T_{2,1}} = \{a_1^\dag a_1^\dag a_1,  a_1^\dag a_1^\dag a_2, \ldots,  a_d^\dag a_d^\dag a_d\}\,,\q \mc{W}_{T_{2,2}} = \{a_1 a_1 a_1^\dag,  a_1 a_1 a_2^\dag, \ldots, a_d a_d a_d^\dag\}\,.
\end{align*}
\normalsize
With the same notation as in \eqref{eq:defl21}, we define
\small
\begin{align} \label{eq:setl12t}
     \mc{L}_+^{(1,2),T} :=  \Big\{\{h_{\gamma \d}\}_{\gamma < \d}   \,;\  &\sum_{\gamma < \d}h_{\gamma \d} = \tr \big(Y\ff^{\mc{W}_1}\big) + \sum_{\gamma < \d} g_{\gamma \d} + \tr \big(Y_{T_1} \ff^{\mc{W}_{T_{1,1}}}\big) + \tr\big(\overline{Y_{T_1}} \ff^{\mc{W}_{T_{1,2}}}\big) \notag \\
     & +\tr \big(Y_{T_2} \ff^{\mc{W}_{T_{2,1}}}\big) + \tr\big(\overline{Y_{T_2}} \ff^{\mc{W}_{T_{2,2}}}\big) 
     \,,\  \forall Y,\, Y_{T_1}, Y_{T_2} \ge 0, \, \mc{A}^{{\rm sa}}_{V_{\gamma \d}}
\ni g_{\gamma \d } \ge 0  \Big\}\,,
\end{align}
\normalsize
where $\ff^{\sss \mc{W}_{T_{i,j}}}$, for $i,j = 1,2$, is defined by \eqref{def:matrix_list}, and $\overline{Y_{T_i}}$, for $i = 1,2$, is the conjugate matrix of $Y_{T_i}$. Then, by replacing $\mc{L}_+^{\sss (1,2)}$ in \eqref{scheme2} with $\mc{L}_+^{\sss (1,2),T}$ in \eqref{eq:setl12t}, we readily have a tighter relaxation enhanced by the RDM $T_1$ and $T_2$ conditions, denoted by $E_0^{\sss (1,2),T}[H]$. Similarly to \eqref{eq:dualscheme2}, we also have 
\small
\begin{align} \label{eq:t1embedding}
     E_0^{(1,2),T}[H] 
    & = \sup_{\mycom{0\le g_{\gamma \d} \in\mc{A}^{{\rm sa}}_{V_{\gamma \d}}, \lad \in \R}{P_{T_{i,j}}\ \text{in}\ \eqref{eq:poly1}\ \text{and}\ \eqref{eq:poly2}   }} \bigg\{\lad \,;
    H - \lad  = 1\text{-SOS} + \sum_{\gamma < \d} g_{\gamma \d}  + \sum_{i,j = 1}^2 P_{T_{i,j}}^\dag P_{T_{i,j}} \bigg\}\,.
\end{align}
\normalsize
Now, recalling the definition \eqref{2rdmt1t2} of $E_0^{\sss (2,3),{\rm RDM}}$, we see that the relaxation $E_0^{\sss (1,2),T}$ is at least as tight as the larger one in $E_0^{\sss (2,3),{\rm RDM}}$ and $E_0^{\sss (1,2)}$, namely, 
\begin{align} \label{eq:rela_rdm}
\max\big\{E_0^{(2,3),{\rm RDM}}[H],  E_0^{(1,2)}[H]  \big\}  \le E_0^{(1,2),T}[H] \le E_0[H]\,.
\end{align}

\medskip 

We finally investigate the relations with the Anderson bound \cite{anderson1951limits,wittmann1993bounds}. Without loss of generality, we fix a partition \eqref{eq:index_cluster_fermi} and a grouping \eqref{eq:group_v}, and we assume that the fermionic Hamiltonian $H$ can be split as 
\begin{align} \label{def:anderson}
    H = \sum_{\gamma < \d} \h{H}_{\gamma \d}\,,
\end{align}
where $\h{H}_{\gamma \d} \in \mc{A}^{{\rm sa}}_{V_{\gamma \d}}$ are local Hamiltonians given in \eqref{eq:split_hamil} with $X_{C_i} = - H_{C_i}$. Then the Anderson lower bound is simply defined by
\begin{align*}
E_0^{{\rm Ander}}[H] : = \sum_{\gamma < \d} E_0[\h{H}_{\gamma \d}] \le E_0[H]\,.
\end{align*}
We claim that $E_0^{\sss (k)}$ for $k \ge 2$ can be reformulated as $\sum_{\gamma < \d} E_0[\h{H}_{\gamma \d}^{{\rm eff}}]$ but for some effective Hamiltonians $\{\h{H}_{\gamma \d}^{{\rm eff}}\}$ optimized in a certain class.
We consider the Lagrangian formulation of $E_0^{\sss (k)}$ from \eqref{scheme_general}: 
\small
\begin{align} \label{scheme_lag}
         E_0^{(k)}[H] 
        =  \inf_{\mycom{\vr^{C_i} \in \mc{D}(\mc{A}_{C_i})}{0 \le l^{\gamma \d} \in (\mc{A}_{V_{\gamma \d}}^{{\rm sa}})'}} \sup_{\mycom{X_{C_i} \in \mc{A}^{{\rm sa}}_{C_i}, \{h_{\gamma \d}\}_{\gamma < \d} \in \mc{L}_+^{(k)}}{X_{\gamma \d, \gamma' \d'} + X_{\gamma' \d'
        , \gamma \d} = 0}}   \bigg(& \sum_i \vr^{C_i}\big(X_{C_i}  + H_{C_i}\big) \notag \\ & + \sum_{\gamma < \d} l^{\gamma \d} \Big(H_{\gamma \d}- h_{\gamma \d}  + \sum_{\mycom{\gamma' < \d'}{(\gamma,\d) \neq (\gamma',\d')}} X_{\gamma \d, \gamma' \d'}\Big)  
        \bigg)\,.
\end{align}
\normalsize
Thanks to the supremum in \eqref{scheme_lag} over the self-adjoint operators $X_{\sss C_i}$ and $X_{\sss \gamma \d, \gamma' \d'}$, we find that the optimization variables $\vr^{\sss C_i}$ and $l^{\sss \gamma \d}$ satisfy the consistency \eqref{eq:consis_const} and $l^{\sss \gamma \d}|_{\sss C_i} = \vr^{\sss C_i} $ for $i \in V_{\sss \gamma \d}$. 
 We hence obtain 
 \small
 \begin{align} \label{scheme_lag_2}
         E_0^{(k)}[H] 
      &  =  \inf_{l^{\gamma \d} \in \mc{D}(\mc{A}_{V_{\gamma \d}})} \sup_{\mycom{\{h_{\gamma \d}\}_{\gamma < \d} \in \mc{L}_+^{(k)}}{X_{\gamma \d, \gamma' \d'} + X_{\gamma' \d'
        , \gamma \d} = 0}}  \sum_{\gamma < \d} l^{\gamma \d} \Big(\h{H}_{\gamma \d}- h_{\gamma \d}  + \sum_{\mycom{\gamma' < \d'}{(\gamma,\d) \neq (\gamma',\d')}} X_{\gamma \d, \gamma' \d'}\Big) \notag \\
        & =  \sup_{\mycom{\{h_{\gamma \d}\}_{\gamma < \d} \in \mc{L}_+^{(k)}}{X_{\gamma \d, \gamma' \d'} + X_{\gamma' \d'
        , \gamma \d} = 0}}  \sum_{\gamma < \d} E_0 \big[\h{H}^{{\rm eff}}_{\gamma \d}\big]\,,
\end{align}
\normalsize
with the effective Hamiltonian on the cluster $V_{\gamma \d}$:
\begin{align} \label{auxclaim1}
\h{H}^{{\rm eff}}_{\gamma \d}: = \h{H}_{\gamma \d}- h_{\gamma \d}  + \sum_{\mycom{\gamma' < \d'}{(\gamma,\d) \neq (\gamma',\d')}} X_{\gamma \d, \gamma' \d'},
\end{align}
where $\h{H}_{\gamma \d}$ is defined as in \eqref{def:anderson}. Then, there holds, by letting $h_{\gamma \d} = 0$ and $X_{\gamma \d, \gamma' \d'} = 0$, 
\begin{align} \label{auxclaim2}
     E_0^{(k)}[H] \ge E_0^{{\rm Ander}}[H]\,.
\end{align}

\begin{remark} \label{rem:inter_eff} The formulation \eqref{scheme_lag_2} naturally admits a quantum embedding interpretation as follows: the fragments are given by the clusters $\{V_{\gamma \d}\}$ with the exact diagonalization as the high-level local calculation; the correction term  $- h_{\gamma \d}  + \sum_{\sss \gamma' < \d',(\gamma,\d) \neq (\gamma',\d')} X_{\gamma \d, \gamma' \d'}$ in the effective Hamiltonian \eqref{auxclaim1} describes the effect of the environment on the fragment $V_{\gamma \d}$; these local energies $E_0 [\h{H}^{{\rm eff}}_{\gamma \d}]$ are glued together to predict the global one $E_0[H]$ by the constraints $\{h_{\gamma \d}\}_{\gamma < \d} \in \mc{L}_+^{\sss (k)}$ and the communication variables $X_{\gamma \d, \gamma'\d'}$ between fragments, with the SDP as the low-level global calculation. 
\end{remark}

\begin{remark}\label{rem:inter_eff22}
    From \eqref{eq:joint_rep}  and \eqref{scheme_lag_2}, it is easy to see that the optimal dual variables ${l^{\gamma \d} \in \mc{D}(\mc{A}_{V_{\gamma \d}})}$ are nothing else than  the (optimal) relaxed marginals, which satisfy the KKT conditions by optimality, and  achieve the ground state energies $E_0[\h{H}^{\rm eff}_{\gamma \d}]$ of effective Hamiltonians by the complementary slackness. Noting from numerical experiments that the lowest energy of $\h{H}^{\rm eff}_{\gamma \d}$ is highly degenerate, to find the relaxed marginals, one has to use the optimality conditions instead of simply solving $E_0[\h{H}^{\rm eff}_{\gamma \d}]$; see also Remark \ref{rem:imple_ek}. 
\end{remark}

\begin{remark} \label{rem:simple_eff} 
The same arguments as above can be applied to the relaxations $E^{\sss (1,2)}_0$ in \eqref{scheme2} and $E^{\sss (1,2)'}_0$ in \eqref{scheme3}. In particular, $E_0^{\sss (1,2)'}[H]$ has a very simple formulation:
\begin{align} \label{scheme3_2}
       E_0^{(1,2)'}[H] 
        & =  \sup_{Y \ge 0}  \sum_{\gamma < \d} E_0 \big[\h{H}_{\gamma \d}- \w{H}_{\gamma \d}\big].
\end{align} 
\end{remark}

\subsection{Extensions to quantum spin system} \label{subsec:spin}
 We extend the above discussions to the quantum spin system. We start with the basic setup. We consider a system of $N$ qubits (spins), the state space of which is given by $\C^{D} \simeq \otimes_{i = 1}^N \C^{2}$ with $D = 2^N$. We denote by $\mc{Q}_S$ the Hilbert space $\otimes_{i \in S} \C^{2}$ for a subset $S$ of $[N]$ and by $\mc{B}(Q)$ (resp., $\mc{H}(Q)$) the space of bounded (resp., Hermitian) operators on a Hilbert space $Q$. Moreover, for an operator $A \in \mc{B}(Q_{S})$,
we define the operator $\h{A} \in \mc{B}(\C^D)$ by tensoring $A$ with the identity on the remaining sites $k \in [N] \backslash S$. For simplicity, we focus on the following Hamiltonian on $\C^D$ with two-body interactions: 
\begin{equation} \label{def:hamitonian}
    H = \sum_{1 \le i \le N} \h{H}_i + \sum_{1 \le i < j \le N} \h{H}_{ij}\,,
\end{equation}
the ground-state energy of which is defined by 
\begin{align} \label{model_1}
    E_0[H] = \inf_{\rho \in \mc{D}(\C^D)} \tr(H \rho)\,,
\end{align}
where $H_i \in \mc{H}(\C^{2})$ and $H_{ij} \in \mc{H}(\C^{2} \otimes \C^{2})$ are local Hamiltonians, and 
$\mc{D}(\C^D): = \{\rho \in \mc{H}(\C^D)\,;\ \rho \ge 0 \,,\ \tr(\rho) = 1\}$ is the convex set of density matrices on $\C^D$. 

Let $\si^s$ for $s = o,x,y,z$ be the Pauli matrices:
\begin{align*}
\si^{o} = I\,,\q   \si^{x} =  \mm 0 & 1 \\ 1 & 0  \nn\,, \q  \si^{y} = \mm 0 & -i \\ i & 0  \nn\,, \q  \si^{z} =  \mm 1 & 0 \\ 0 & -1  \nn\,.
\end{align*} 
For $i \in [N]$ and $s \in \{o,x,y,z\}$, we define the generalized Pauli matrices $\si_i^{s}: = \si^{s} \otimes {\rm id}_{[N] \backslash i}$, which satisfy
the commutation relations:
\begin{align} \label{eq:commu_rela}
    [\si_j^{s}, \si_l^{t}] = i \d_{jl} \ep_{str} \si_j^{r}\,, \q  j,l\in [N]\,,\ s,t,r \in \{x,y,z\}\,,
\end{align}
where $\ep_{str}$ is the Levi-Civita symbol (view $x,y,z$ as $1,2,3$). The $*$-algebra generated by $\si_i^{s}$ gives all the square matrices on $\C^D$ with the basis:
\begin{align} \label{def:basis_pauli}
    \si_i^{s_1} \si_i^{s_2} \cdots \si_i^{s_N}\,,\q s_i \in \{o,x,y,z\}\,. 
\end{align}
Thus, the $2$-local Hamiltonian \eqref{def:hamitonian} can be written as a polynomial in $\{\si_i^{s}\}$ with each monomial depending only on two sites $i,j \in [N]$, 
namely, 
\begin{align} \label{hamil_poly}
    H = \sum_{0 \le i < j \le [N]} \sum_{s, t \in \{o,x,y,z\}} c_{ij}^{st} \si_i^{s} \si_j^{t}\,,
\end{align}
for some coefficients $c_{ij}^{st}$. 
It is easy to see that the abstract notions of quantum state and marginal in Section \ref{sec:basic} for the fermionic case apply to the spin case as well.  
In detail, we similarly define the quantum states by the positive linear functionals on $\mc{B}(\C^D)$ with trace one. Then the Riesz representation allows us to identify the density matrices $\rho$ with the  states $\vr$ by $\vr(\dd):= \tr (\rho\, \dd)$. For consistency of exposition,  we still use $\mc{A}_S$ to denote the subalgebra on the cluster $S$:
\begin{align} \label{def:paulialgebra}
\mc{A}_S := \big\l \{\si_i^{s}\,;\ i \in S\,,\ s \in \{o,x,y,z\}\} \big\r\,,
\end{align}
and use $\mc{A}_S^{{\rm sa}}$ for its self-adjoint elements. It is clear that $\mc{A}_S = \mc{B}(Q_S)$ and $\mc{A}_S^{{\rm sa}} = \mc{H}(Q_S)$, and that the quantum 
marginal $\vr^S := \vr|_{\mc{A}_{S}}$ gives the partial trace  $\rho^{S}: = \tr_{[N]\backslash S} (\rho)$ by $\vr^S(A) = \tr(\rho^S A)$ for $A \in \mc{A}_S$.  

From above observations, we conclude that the only difference between the problems \eqref{subtarget} and \eqref{model_1} lies in the generators and commutation relations of the underlying algebras (\eqref{eq:commu} for the fermionic system and \eqref{eq:commu_rela} for the spin one). As a result, the dual formulation \eqref{eq:dual_ground2} and the variational embedding hierarchy \eqref{eq:relaxation} (including its variants \eqref{scheme2} and \eqref{scheme3}) remain applicable for spin Hamiltonians but with the  algebra generated by the Pauli matrices with \eqref{eq:commu_rela}. Furthermore, the relationships with other variational methods \cite{anderson1951limits,lin2022variational,haim2020variational} can be established similarly to those in Section \ref{subsec:connection}
without any difficulty, although we want to remark that the analog of the RDM method for the spin system seems not well-developed yet and we only note the work \cite{haim2020variational} that follows the same logic as that of the RDM method.

\begin{remark}[(\emph{Mapping to the local fermionic Hamiltonian})] Alternatively, instead of analyzing the quantum spin system with the algebra \eqref{def:paulialgebra}, one can map the local spin Hamiltonian to the local fermionic one via the Schwinger representation \cite{verstraete2005mapping,liu2007quantum,wei2010interacting}, which allows us to directly apply the results for the fermionic system to the spin case. The key idea of the representation is to map each qubit at site $i$ to a spinful fermion with $a_{i,\uparrow}$ and $a_{i,\down}$. To be precise, 
we consider a system of $N$ spinful fermions with $N$ orbitals and associate a $N$-qubit state with a fermionic state by 
    \begin{align*}
    \ket{z_1}\otimes \ldots \otimes \ket{z_N} \mapsto (a_{1,\up}^\dag)^{1 - z_1} (a_{1,\down}^\dag)^{z_1} \ldots (a_{N,\up}^\dag)^{1 - z_N} (a_{N,\down}^\dag)^{z_N} \ket{\Omega},\q z_i \in \{0,1\}\,.
\end{align*}
We then define, for $i \in [N]$, 
\begin{align*}
    S_i^{x} := a_{i,\up}^\dag a_{i,\down} + a_{i,\down}^\dag a_{i,\up}\,, \q  S_i^{y} :=   i(a_{i,\down}^\dag a_{i,\up} - a_{i,\up}^\dag a_{i,\down})\,, \q S_i^{z} := n_{i,\up} - n_{i,\down}\,.
\end{align*}
It is easy to check that $\{S_i^{s}\}_{s \in \{x, y, z\}}$ satisfies the Pauli commutation relation \eqref{eq:commu_rela}. Hence, we can identify $S_i^{s}$ with the Pauli matrix $\si_i^{s}$ and map the local spin Hamiltonian \eqref{hamil_poly} to the following fermionic one: \begin{align*} 
    H = \sum_{0 \le i < j \le [N]} \sum_{s, t \in \{o,x,y,z\}} c_{ij}^{st} S_i^{s} S_j^{t}\,,
\end{align*}
subject to the constraint: for $i \in [N]$, $n_{i,\up} + n_{i,\down} = 1$. 
\end{remark}

We end this section with some prototypical quantum spin-$1/2$ Hamiltonians that will be used in our numerical simulations. We mainly focus on the transverse-field Ising (TFI) model with $N$ spins:
\begin{align} \label{def:tfimodel}
    H = - \sum_i h_i \si_i^x - \sum_{i \sim j} J_{ij} \si_i^z \si_j^z\,, 
\end{align}
where $J_{ij} \in \R$ is the interaction strength between adjacent spins, 
and $h_i \in \R$ is the 
strength of the transverse magnetic fields. 
When $h_i$ and $J_{ij}$ are i.i.d. Gaussian
with zero mean $\mathbb{E}[J_{ij}] = \mathbb{E}[h_i] = 0$ and variance $\mathbb{E}[h_i^2] = h_0^2$ and $\mathbb{E}[J_{ij}^2] = J_0^2/N$, where $h_0$ and $J_0$ are constants,
the Hamiltonian \eqref{def:tfimodel} on a complete graph gives a paradigmatic disordered model: the quantum Sherrington-Kirkpatrick (SK) model \cite{ishii1985effect,suzuki2012quantum}. 
We will also consider the quantum Heisenberg XXZ model on a general graph: 
 \begin{align} \label{def:xxzmodel}
H = \sum_{i \sim j}  \si_i^x \si_j^x + \si_i^y \si_j^y + J_z \si_i^z \si_j^z\,,
\end{align}
with real coefficient $J_z \in \R$. We refer interested readers to \cite{cubitt2016complexity} for more quantum spin models.


\begin{remark}[\emph{(Perturbation theory)}] Very recently, Hastings \cite{hastings2022perturbation,hastings2023field} explored the connections between the sum-of-squares hierarchy and the perturbation theory, as well as the quantum field theory, for spin
and fermion systems. It would be interesting to establish similar results in our setup. Some preliminary results can follow directly from Section \ref{subsec:connection} and \cite{hastings2022perturbation,hastings2023field}. For instance, we consider the TFI model \eqref{def:tfimodel} with $|J_{ij}| \ll 1$ for all $i,j$.  By \eqref{eq:compare} and the result in \cite[Section I]{hastings2022perturbation}, we conclude that the SDP relaxation $E^{\sss (1)}_0[H]$ can at least produce the asymptotics of $E_0[H]$ in $J: = (J_{ij})$ up to the third order, i.e., $E^{\sss (1)}_0[H] = E_0[H] + O(\norm{J}^4)$.

\end{remark}

\section{Variational embedding with optimized clusters} \label{sec:multiembedding}

As seen in Remarks \ref{rem1} and \ref{rem:sublevel}, enlarging clusters can systematically improve the accuracy of variational embedding relaxation, but it could be expansive for large clusters due to the exponential scaling 
of computational cost with respect to the cluster size. To balance the computational cost and accuracy, it is natural to design a strategy to optimally choose the clusters of orbitals or sites. This section is devoted to this purpose. 

Our strategy is motivated by quantum information theory. One of the fundamental properties of a multipartite quantum system is entanglement which refers to the non-classical correlations 
between multiple sites. Quantifying the entanglement is crucial for understanding the behavior of many-body systems and many related numerical algorithms, for example, the DMRG. Since a quantum embedding method is usually very accurate at a local scale, one may expect that its efficiency
can be maximized by the clusters such that the correlation or entanglement distribution of the ground state is concentrated inside them.


\subsection{Quantum entanglement and correlation} \label{sec:basic_corre}
We will briefly review the basic concepts of entanglement and correlation measures for quantum systems. 
Let us start with the simpler spin case. We consider a bipartite system $Q_{AB} = Q_A \otimes Q_B$ with the observable algebra $\mc{A}^{{\rm sa}} = \mc{A}_{A}^{{\rm sa}}  \otimes \mc{A}_{B}^{{\rm sa}}$, where $Q_{A/B}$ and $\mc{A}_{A/B}^{{\rm sa}} = \mc{H}(Q_{A/B})$ are the local Hilbert space and observables on the subsystem $A/B$. 
\begin{definition} \label{def:entanglement}
A bipartite state $\rho^{AB} \in \mc{D}(\mc{Q}_{AB})$ 
is \emph{uncorrelated} with respect to subsystems $A$ and $B$ if 
\begin{align*}
    \tr\big(\rho^{AB} O_A \otimes O_B\big) = \tr\big(\rho^A O_A\big) \tr\big(\rho^B O_B\big)\,,\q  \forall\, O_{A/B} \in \mc{A}_{A/B}^{{\rm sa}}\,,
\end{align*}
equivalently, $\rho^{AB} = \rho^A \otimes \rho^B$ is a \emph{product state}, where $\rho^{A/B}$ is the marginal of $\rho^{AB}$.  We denote by $\mc{D}_0(Q_{AB})$ the set of uncorrelated states, and by $\mc{D}_{{\rm sep}}(Q_{AB})$ the convex hull of $\mc{D}_0$. We say  $\rho^{AB}$ is \emph{separable} for subsystems $A$ and $B$ if $\rho^{AB}  \in \mc{D}_{{\rm sep}}$; otherwise, we say it is \emph{entangled}. 
\end{definition}

Note that the concepts of quantum correlation and entanglement depend on the partition of the system.
A natural question is 
how to determine whether or not a given quantum state is correlated or entangled. 
Efficient criteria are known for this task, such as the Schmidt rank criterion for pure states and the positive partial transpose (PPT) criterion for mixed states; see \cite{khatri2020principles} for details.

For our purpose, we are more interested in measures 
quantifying the correlation and entanglement in a state.
Recall that $E(A,B)_{\rho}: \mc{D}(Q_{AB}) \to \R$ is an entanglement measure if there holds 
\begin{align} \label{def:ent_mea}
    E(A,B)_\rho \ge  E(A', B')_{w}\,,
\end{align}
for any bipartite state $\rho^{AB}$ and local operations and classical communication (LOCC) channel $\Phi_{AB \to A'B'}$ from $\mc{D}(Q_{AB})$ to $\mc{D}(Q_{A'B'})$, where $\ww = \Phi_{AB \to A'B'}(\rho)$.  In this work, for simplicity, we only consider the divergence-based measure: 
 \begin{align} \label{entan_es}
        E_S(A,B)_\rho := \inf_{\si \in \mc{D}_{{\rm sep}}(Q_{AB})}  S(\rho^{AB} || \si)\,,
    \end{align}
where $S(\rho || \si) = \tr\rho ( \log\rho - \log\si)$ is the quantum relative entropy. 
Other entanglement measures satisfying \eqref{def:ent_mea}
include the entanglement of formation, the logarithmic negativity, etc. If we restrict the constraint in \eqref{entan_es} to uncorrelated states, we obtain the \emph{mutual information} that quantifies the total correlation of the system, that is,
\begin{align} \label{def:total_corre}
 \inf_{\si \in \mc{D}_{0}(Q_{AB})}  S(\rho^{AB} || \si) = I(A,B)_\rho := S(\rho^A) + S(\rho^B) - S(\rho^{AB})\,,
\end{align}
where the infimum is attained at $\rho^A \otimes \rho^B$ and $S(\rho) = - \tr(\rho \log \rho)$ is the von Neumann entropy. We remark that $I(A,B)_\rho$ is not an entanglement measure, since it may increase under some LOCC channels. See \cite[Section 5]{khatri2020principles} for a complete review. 

\begin{remark}
      There is another class of measures quantifying the quantum correlation, known as quantum discord\cite{henderson2001classical,ollivier2001quantum,modi2010unified}, which we do not consider in this work. 
    It is a weaker concept than entanglement in the sense that the unentangled states can still present non-zero quantum discord. 
\end{remark}

We next consider the case of indistinguishable fermions, and define the fermionic entanglement and correlation between the clusters of orbitals, following \cite{banuls2007entanglement,ding2020concept}.
Let
\begin{align*}
    [d] = A \cup B \q \text{with}\q  A := \{1,\ldots,m\}\,,\ B := \{m+1,\ldots,d\}\,,
\end{align*}
and $\mc{F}_d = \mc{F}_A \otimes \mc{F}_B$ be a 
bipartite Fock space. The observables $\mc{A}_{d}^{{\rm sa}} = \mc{A}_{A}^{{\rm sa}} \otimes \mc{A}_{B}^{{\rm sa}}$ are self-adjoint operators generated by $\{a_i, a_i^\dag\}$. The analog of Definition \ref{def:entanglement} in the fermionic case is as follows. 

\begin{definition} \label{def:fer_eng}
   For a  bipartite state $\vr^{AB} \in \mc{D}(\mc{A}_d)$, we say it is uncorrelated if 
    \begin{align}\label{eq:product_fermi}
    \vr^{AB}(O_A O_B) = \vr^A(O_A) \vr^B(O_B)\,,\q  \forall\, O_{A/B} \in \mc{A}^{{\rm sa}}_{A/B}\,,
    \end{align}
  and it is separable if it is a convex combination of uncorrelated states; otherwise, it is entangled. Moreover, $\mc{D}_0$ and $\mc{D}_{{\rm sep}} := {\rm Conv}(\mc{D}_0)$ still denote the uncorrelated and separable states, respectively. 
\end{definition}

With Definition \ref{def:fer_eng} above, we can similarly define the entanglement and correlation of a fermionic state $\vr^{AB}$ by 
its relative entropy to the sets $\mc{D}_{{\rm sep}}$ and $\mc{D}_0$, respectively. However, their numerical computations are a bit subtle, as we need to use the Jordan-Wigner transformation (JWT) to map the fermionic operators to spin ones, which is defined by, for any $i \in [d]$, 
\begin{align} \label{def:jwt}
   \mc{J}_d(a_i^\dag) = \otimes^{i-1}\si^{z} \otimes \mm 0 & 0 \\ 1 & 0 \nn \otimes^{d-i} I\,.
\end{align}
Thanks to JWT, the density matrix $\rho \in \mc{D}(\C^{2^d})$ associated with the state $\vr$ is given
by $\rho = \mc{J}_d^*(\vr)$, i.e.,
\begin{align} \label{def:density_matrix_dual}
    \vr(O) = \tr[\rho \mc{J}_d(O)]\,, \q \forall O \in \mc{A}_d\,.
\end{align}
A drawback of the JWT is that it does not keep the locality of the system, which means that $\mc{J}_d^*(\vr^{AB}) \in \mc{D}(Q_{AB})$ for an uncorrelated state $\vr^{AB}$ may not be a product state on $Q_{AB} = Q_A \otimes Q_B$. However, as shown in \cite[A.6]{banuls2007entanglement}, we can still have 
the equivalence between the uncorrelated states and the product states, if we restrict to the physical fermionic states. 

Indeed, the physically meaningful observables satisfy the so-called \emph{superselection rules}. To be precise,  we recall the parity operator $P := \prod_i (1 - 2 \h{n}_i)$ on the Fock space $\mc{F}_d$, which has eigenvalues $\pm 
1$ with associated eigenspaces $\mc{F}_{e/o}$ consisting of states with even/odd number of fermions. The corresponding  projections from $\mc{F}_d$ to $\mc{F}_{e/o}$ are denoted by $P_{e/o}$. We introduce 
the physical observable algebra by
\begin{align} \label{def:physical_ob}
    \mc{A}_d^{{\rm sa},\pi} := \{O \in \mc{A}_d^{{\rm sa}}\,;\ [O,P] = 0\}\,.
\end{align}
Note that $O \in \mc{A}_d^{{\rm sa},\pi}$ if and only if $O = P_e O P_e + P_o O P_o$ holds, since the operators $O$ and $P$ are simultaneously diagonalizable.  
The physical states $\mc{D}^\pi(\mc{A}_d)$ are then defined by those characterized by the physical observables,  that is, $\vr \in \mc{D}^\pi(\mc{A}_d)$ if it vanishes on the non-physical observables, i.e.,
\begin{align} \label{def:physical_state}
  \vr(P_o O P_e + P_e O P_o) = 0\,,\q \forall O \in \mc{A}_d^{{\rm sa}}\,.
\end{align}

We claim that $\mc{J}_d^*(\vr^{AB})$ for a uncorrelated physical state $\vr^{AB} \in \mc{D}^\pi(\mc{A}_d) \cap \mc{D}_0(\mc{A}_d)$ can be written as 
\begin{align} \label{lem:prod}
\rho^{AB}:= \mc{J}_d^*(\vr^{AB}) = \rho^A \otimes \rho^B,  \q \text{for some}\ \rho^{A/B} \in \mc{D}(Q_{A/B}).  
\end{align}
We provide a sketch of proof for completeness.
Noting that $P_e = (P + I)/2$ and $P_o = (I - P)/2$, we have that $O \in \mc{A}_d^{{\rm sa},\pi}$ is equivalent to $O = P O P$.  
By a direct computation with \eqref{eq:commu}, we see that $1 - 2 \h{n}_i$ anti-commutes with $a_i, a_i^\dagger$ and commutes with $a_j, a_j^\dagger$ 
for $j \neq i$, which, along with the property of the parity operator $P^2 = I$, implies $POP = O_e - O_o$ for a fermionic polynomial $O = O_e + O_o \in \mc{A}_d$, where 
$O_e$ and $O_o$ are the even and odd parts of  $O $, respectively. Hence, $O \in \mc{A}_d^{{\rm sa},\pi}$ if and only if $O$ is even. Then, by Definition \ref{def:fer_eng},  it suffices 
to consider the physical observables $O_A \in \mc{A}^{{\rm sa}, \pi}_{A}$ and $O_B \in \mc{A}^{{\rm sa}, \pi}_{B}$ in 
\eqref{eq:product_fermi}, if $\vr^{AB}$ is a physical state. Let $\mc{J}_{A/B}$ be the local JWT on the algebra $\mc{A}_{A/B}$ and 
write $\rho^{A/B} = \mc{J}_{A/B}^* (\vr^{A/B})$.  Since $O_A$ and $O_B$ are even, we have $\mc{J}_d(O_A) = \w{O}_A \otimes I$ and $\mc{J}_d(O_B) = I \otimes \w{O}_B$ with $\w{O}_{A/B} = \mc{J}_{A/B}(O_{A/B})$, and there holds
\begin{align} \label{auxeq:prod}
    \tr\big(\rho^{AB} \w{O}_A \otimes \w{O}_B\big) = \tr\big(\rho^A \w{O}_A \big) \tr\big(\rho^B \w{O}_B\big)\,.
\end{align}
Using the fact $\vr^{AB}(O) = 0$ for the odd observables $O \in \mc{A}_d^{{\rm sa}}$,
it is easy to check that \eqref{auxeq:prod} can hold for any self-adjoint $\w{O}_A \in \mc{H}(Q_A)$ and $\w{O}_B \in \mc{H}(Q_B)$. The claim \eqref{lem:prod} is proved.

The property \eqref{lem:prod} enables us to compute the fermionic entanglement and correlation
for a physical state $\vr^{AB}$ as in \eqref{entan_es} and \eqref{def:total_corre} by JWT:
\begin{align*}
    E_S(A,B)_{\vr} = E_S(A,B)_{\mc{J}_d^*(\vr)}\q \text{and}\q I(A,B)_{\vr} = I(A,B)_{\mc{J}_d^*(\vr)}. 
\end{align*}

\begin{remark}[(\emph{Computation of entanglement})] 
The entanglement measure $E_S(A,B)_\rho$ in \eqref{entan_es} 
is a conceptually simple convex optimization, but the boundary of the feasible set $\mc{D}_{\rm sep}(Q_{AB})$ is hard to characterize, which makes its computation a nontrivial task. Recalling Carath\'{e}odory's theorem in convex analysis, we can parameterize $\mc{D}_{\rm sep}(Q_{AB})$ as follows, 
\small
\begin{align} \label{eq:sep_para}
    \mc{D}_{\rm sep}(Q_{AB}) = \Big\{\sum_{i = 1}^{M} \si^A_i \otimes \si_i^B\,;\ \si_i^{A/B} \ge 0\,,\ \tr\Big(\sum_{i = 1}^{M} \si^A_i \otimes \si_i^B\Big) = 1 \Big\}\,,\q \text{where}\ M = {\rm dim}(Q_{AB})^2\,,
\end{align}
\normalsize
and then minimize $S \big(\rho^{AB} || \sum_i \si^A_i \otimes \si_i^B \big)$ over the set \eqref{eq:sep_para}, which is a non-convex optimization in $\{\si_i^A,\si_i^B\}_{i \in [M]}$.
However, noting that the relative entropy $S(\rho || \si)$ is jointly convex \cite{khatri2020principles}, it could be efficiently solved by an alternating minimization in $\{\si_i^A\}$ and $\{\si_i^B\}$, where each sub-optimization problem is a standard SDP. To avoid the local minima, we run the algorithm repeatedly with random initial states $\{\si_i^A,\si_i^B\}$ and set $E_S(A,B)_\rho$ as the minimal output.
\end{remark}

\begin{remark}[(\emph{Overlapping subsystems $A$ and $B$})]  In the sequel, we may also need to compute the correlation or entanglement of a state on $\mc{Q}_{A\cup B}$ but with $A \cap B \neq \emptyset$. In this case, we modify the definition of $I(A,B)_\rho$ as $I(A,B)_\rho = \max\{I(A,B\backslash A)_\rho, I(B,A\backslash B)_\rho\}$, where $A\backslash B$ and $B \backslash A$ are assumed to be not empty. Similar convention applies to the entanglement $E_S(A,B)_{\rho}$. 
\end{remark}

\subsection{Cluster optimization} \label{sec:adaptive_embedding}
We shall propose several simple strategies, based on quantum information measures, to optimize the cluster selection and tighten the SDP 
hierarchy in Section \ref{sec:sosemb}. For ease of exposition and clarity, in the remaining of this work, 
\begin{itemize}[leftmargin= 8 mm]
    \item we limit our discussion to the Hamiltonian of the form \eqref{hamiltonian_decom} with 
    $C_i = \{i\}$ in \eqref{eq:index_cluster_fermi} and consider the selection of groups $V_\gamma$ in \eqref{eq:group_v}. 
    \item we focus on the relaxation $E_0^{\sss (1,2)}$ in \eqref{scheme2} 
(i.e., the one in \cite{lin2022variational})  and re-denote it by $E_0^{\sss \rm sdp}$ for convenience. Our strategy does not depend on the relaxation scheme we use. In principle, one can replace $E_0^{\sss \rm sdp}$ with any relaxation $E_0^{\sss (k)}$ in the hierarchy. 
\item by the uniform $k \times 1$ clusters, we mean $\{V_1 = \{1,\ldots,k\}, V_2 = \{k+1,\ldots,2k\},\ldots\}$.
\end{itemize}

\begin{figure}[!htbp]
  \centering
  \includegraphics[width=0.65\textwidth]{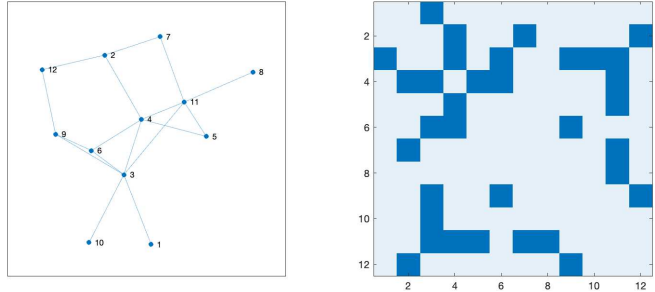}
  \caption{An instance of Erd\"{o}s-R\'{e}nyi graph $G(12,0.3)$ (left) and its adjacency matrix (right). 
  }
  \label{fig:graph1}
\end{figure}

As mentioned above, the main idea is to reduce the relaxation errors by choosing a partition of sites such that the ground 
state of the Hamiltonian is strongly correlated inside clusters where the embedding schemes are highly accurate, while it is weakly correlated between clusters. Let us first consider a toy TFI model \eqref{def:tfimodel} with the underlying graph generated by Erd\"{o}s–R\'{e}nyi model 
$G(N,p)$ with $N = 12$ and $p = 0.3$ (see Figure \ref{fig:graph1}), to show that the correlations of the ground state can indeed help to optimize the cluster choice.
 
We solve $E_0[H]$ with the associated ground state $\rho$ (which turns out to be unique) for the toy TFI model by exact 
diagonalization. We compute the marginals $\rho_{ij}$ of $\rho$ on clusters $\{i,j\}$, and the associated correlation $I(i,j)_\rho$ and 
entanglement $E_S(i,j)_\rho$ by \eqref{entan_es} and \eqref{def:total_corre}, respectively. We expect that if $I(i,j)_\rho \ge I(k,l)_\rho$ (or $E_S(i,j)_\rho \ge E_S(k,l)_\rho$), then grouping $\{i,j\}$ would lead to a tighter SDP relaxation than grouping $\{k,l\}$. If this is indeed the case, then we can regard the correlation $I(i,j)_\rho$ (or entanglement $E_S(i,j)_\rho$)
as an indicator for the clusters in embedding schemes. 

\begin{figure}[!htbp]
  \centering
  \includegraphics[width=0.9\textwidth]{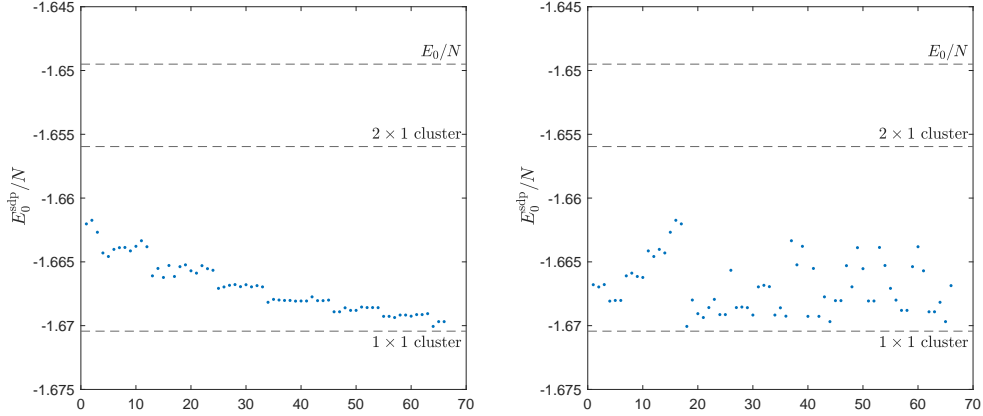}
  \caption{Relaxed energy per site $E_0^{\rm sdp}(i,j)/N$ in $\{(i,j)\}_{i<j}$ ordered by the correlation (left) and entanglement (right).} 
  \label{fig:errordecay}
\end{figure}

To verify this, for each pair $i < j$, we solve the SDP problem  $E_0^{\rm sdp}[H]$ on $1 \times 1$ clusters with a single $2 \times 1$ cluster $\{i,j\}$, i.e., on the non-overlapping clusters 
$\{\{1\},\ldots,\{i-1\},\{i,j\},\{i+1\},\ldots,\{j-1\},\{j+1\},\ldots,\{d\}\}.
$
We denote by $E_0^{\rm sdp}(i,j)$ the relaxed energy for ease of exposition. We order the list of clusters $\{i,j\}$ by the associated correlations $I(i,j)_\rho$: 
\begin{align} \label{def:order_pair}
\{i,j\} < \{k,l\} \q \text{if}\q  I(i,j)_\rho > I(k,l)_\rho\,,
\end{align}
while if $I(i,j)_\rho = I(k,l)_\rho$, we use the dictionary order. Then we plot the relaxed energy per site $E_0^{\rm sdp}/N$ in the ordered list $\{(i,j)\}_{i<j}$ in Figure \ref{fig:errordecay} (left). Similarly, we plot $E_0^{\rm sdp}/N$ in $\{(i,j)\}_{i<j}$ ordered by entanglement (i.e., replacing $I(i,j)_\rho$ in \eqref{def:order_pair} by $E_S(i,j)_\rho$) in Figure \ref{fig:errordecay} (right).
We benchmark against the exact energy per site $E_0/N$ 
and the relaxed ones on uniform $1 \times 1$ clusters and $2 \times 1$ clusters, which are plotted in dashed lines. 

From Figure \ref{fig:errordecay}, it is evident that the relaxation error $(E_0 - E_0^{\rm sdp}(i,j))/N$ almost strictly increases
as the pairwise correlation $I(i,j)_\rho$ decreases, while there is no clear relation between the error and entanglement $E_S(i,j)_\rho$. Thus,
we may conclude that the mutual information \eqref{def:total_corre} is a more suitable indicator for cluster selection than the entanglement \eqref{entan_es}. Moreover, it is remarkable to observe that a single $2 \times 1$ cluster, if chosen correctly, can reduce the error on $1 \times 1$ clusters by around $45\%$, 
 which is much more efficient than tightening the relaxation by simply
 using uniform $2 \times 1$ clusters, in view of the computational cost. 

In practice, the exact marginals are not available. We have to solve the SDP relaxation on $1 \times 1$ 
clusters to extract relaxed marginals $\rho^{\sss \rm sdp}_{ij}$ of the ground state and 
 the associated correlations $I(i,j)_{\sss \rho^{\rm sdp}}$. Note that the marginal errors $\norm{\rho_{ij} - \rho^{\sss \rm sdp}_{ij}}$ are small by numerical experiments, and recall that the mutual information $I(A,B)_\rho$ in \eqref{def:total_corre} is continuous in the state $\rho$ \cite[Section 4.1]{shirokov2021uniform}. We can expect $I(i,j)_{\rho} \approx I(i,j)_{\rho^{\rm sdp}}$ and view $I(i,j)_{\rho^{\rm sdp}}$ as an effective cluster-selection 
indicator. According to the above discussions, we propose Algorithm \ref{alg:optimized_cluster} below to optimize the clusters.

\begin{algorithm}[!htbp]
\caption{Cluster optimization for variational embedding}
\begin{algorithmic}[1]
\Require Hamiltonian $H$; Initial cluster $[d] = \bigcup_i V_i$ with $V_i = \{i\}$;
\Statex \q\q  Integer $k$ with $d \bmod k = 0$; \texttt{keyword} = `overlapping' or `non-overlapping'. 
\Ensure $k \times 1$ overlapping or non-overlapping optimized clusters.
\State Solve $E_0^{\sss \rm sdp}[H]$ on the initial cluster with relaxed marginals $\{\rho_{ij}^{\sss \rm sdp}\}_{i < j}$. 
\State Compute $I(i,j)_{\rho^{\rm sdp}}$ for each $\rho_{ij}^{\sss \rm sdp}$.
\State Order the clusters $V_{ij} := \{i,j\}$ by $I(i,j)_{\rho^{\rm sdp}}$ as in \eqref{def:order_pair}.
\State Define the ordered list by $\texttt{Pair} := \{V^{\sss \rm pair}_l\}_{1\le l \le L}$ with $L = d(d-1)/2$. 
\State Define an empty list for new clusters $\texttt{List} = \{\}$. 
\If{\texttt{keyword} = `overlapping'}
\While{union of clusters in $\texttt{List}$ is a proper subset of $[d]$}
\State Pick the smallest integer $k_c$ such that $\# \big(\bigcup_{1 \le l \le k_c} V_l^{\sss \rm pair} \big) = k$. 
\State Add a new cluster $V^{\sss \rm new}_i := \bigcup_{1 \le l \le k_c} V_l^{\sss \rm pair}$ into $\texttt{List}$. 
\State Remove these $k_c$ clusters from $\texttt{Pair}$ and re-index clusters in $\texttt{Pair}$ from $l = 1$. 
\EndWhile 
\State Remove the possible repeated clusters in $\texttt{List}$ and return $\texttt{List}$. 
\ElsIf{\texttt{keyword} = `non-overlapping'}
\For{$i$ from $1$ to $d/k$} 
\State Initialize a new cluster $V_i^{\sss \rm new}= \{\}$ and an iteration index $\texttt{iter}=1$.  
\Repeat
\If{$\# (V_i^{\sss \rm new} \bigcup V_{\texttt{iter}}^{{\sss \rm pair}}) \le k$ and  $V_{\texttt{iter}}^{\sss {\rm pair}} \bigcap\, (\text{ union of  clusters in $\texttt{List}$}) = \emptyset$} 
\State $V_i^{\sss \rm new} = V_i^{\sss \rm new} \bigcup V_{\texttt{iter}}^{\sss {\rm pair}}$. 
\State Remove $V_{\texttt{iter}}^{\sss {\rm pair}}$ from $\texttt{Pair}$. 
\EndIf 
\State $\texttt{iter} = \texttt{iter} + 1$. 
\Until{ $\# V_i^{\sss {\rm new}} = k$.}
\State Add $V_i^{\sss {\rm new}}$ into $\texttt{List}$. 
\State Re-index clusters in $\texttt{Pair}$ from $l = 1$. 
\EndFor
\State Return $\texttt{List}$. 
\EndIf 
\end{algorithmic} \label{alg:optimized_cluster}
\end{algorithm}

\begin{remark} The main part of Algorithm \ref{alg:optimized_cluster} is the cluster generation, which is a greedy algorithm so that
the existence of optimized clusters can be guaranteed. Moreover, 
its non-overlapping version can be easily adapted for any non-overlapping initial cluster $[d] = \bigcup_i V_i$ with $V_i$ being disjoint.     
\end{remark}

\begin{remark}[(\emph{Implementation of $E_0^{\rm sdp}$ with relaxed marginals})] \label{rem:imple_ek} We first note from \eqref{scheme_general}, \eqref{scheme_lag} and \eqref{scheme_lag_2}, as well as Remark \ref{rem:simple_eff} that the SDP $E_0^{\rm sdp}$ (i.e., $E^{(1,2)}_0$ in \eqref{scheme2}) can be written as 
\small
\begin{align} \label{auxeqscheme}
         E_0^{\rm sdp}[H] 
      =  \sup_{\mycom{\lad_{\gamma \d} \in \R,\ Y \ge 0}{X_{\gamma \d, \gamma' \d'} + X_{\gamma' \d'
        , \gamma \d} = 0}} \bigg\{\sum_{\gamma <  \d} \lad_{\gamma \d}\,; \ \h{H}_{\gamma \d}- \w{H}_{\gamma \d}  + \sum_{\mycom{\gamma' < \d'}{(\gamma,\d) \neq (\gamma',\d')}} X_{\gamma \d, \gamma' \d'} - \lad_{\gamma \d} \ge 0 \bigg\}\,,
\end{align}
\normalsize
with $\h{H}_{\gamma \d}$ and $\w{H}_{\gamma \d}$ given in \eqref{def:anderson} and \eqref{eq:blocksos_2}, respectively. Then, recalling Remark \ref{rem:inter_eff22}, the relaxed marginals are the optimal dual variables for the inequality constraints in \eqref{auxeqscheme}. Therefore, Step 1 in Algorithm \ref{alg:optimized_cluster} (on a general cluster) can be realized by either solving the formulation in \cite[(2.5)--(2.9)]{lin2022variational} or the one in \eqref{auxeqscheme} with some primal-dual algorithm. In this work, we adopt the second choice with \eqref{auxeqscheme}, the implementation of which is uniform for the spin and fermionic systems. 
For the spin case, $\h{H}_{\gamma \d}- \w{H}_{\gamma \d}  + \sum_{\sss \gamma' < \d',\ (\gamma,\d) \neq (\gamma',\d')} X_{\sss \gamma \d, \gamma' \d'} - \lad_{\gamma \d}$ is a Pauli polynomial spanned by \eqref{def:basis_pauli}; for the fermionic case, it is a Clifford polynomial spanned by \eqref{def:basis_cli}. We empathize that for the latter case, the local JWT on $\mc{A}_{V_{\gamma \d}}$ defined as in \eqref{def:jwt} is necessary for mapping the fermionic operator to a matrix. 
\end{remark}

We test the non-overlapping Algorithm \ref{alg:optimized_cluster} with $k = 2,3$ for the toy TFI model as above to generate the optimized $2 \times 1$ and $3 \times 1$ clusters. Then we solve the corresponding SDP problem $E_0^{{\rm sdp}}$ and present the error $(E_0 - E_0^{\rm sdp})/N$ in Table \ref{table:toy_error} below. We observe that compared to the simple uniform partition, the optimized cluster can significantly reduce the relaxation error but with almost the same computational cost (although, to obtain optimized clusters, we have to solve an additional $E_0^{{\rm sdp}}$ on $1 \times 1$ clusters). 

\begin{table}[!htbp]
     \caption{SDP errors with the uniform non-overlapping $2 \times 1$ clusters and $3 \times 1$ clusters, and the optimized $2 \times 1$ clusters: $\{ \{3,4\}, \{6,9\},  \{5,11\},\{2,7\}, \{10,12\},  \{1,8\}\}$ and $3 \times 1$ clusters: $\{\{3,4,11\}, \{6,9,12\}, \{2,5,7\}, \{1,8,10\}\}$.}
     \centering 
   \begin{tabular}{|c|c|c|}
\hline 
Error per site & $2\times1$ cluster & $3\times1$ cluster \tabularnewline
\hline 
uniform & $6.460 \times 10^{-3} $ & $2.308 \times 10^{-3}$ \tabularnewline
\hline 
optimized & $4.283 \times 10^{-3} $ & $8.715 \times 10^{-4}$ \tabularnewline
\hline 
\end{tabular}
    \label{table:toy_error} 
\end{table}

\begin{figure}[htbp]
  \centering
  \includegraphics[width=0.9\textwidth]{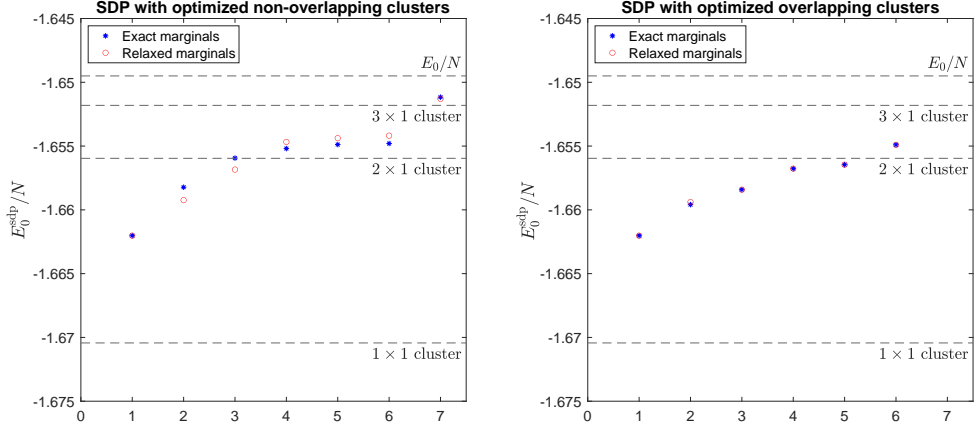}
  \caption{SDP relaxations with sequentially grouped optimized clusters from Algorithm \ref{alg:optimized_cluster}. The dashed lines are for the exact energy per site and the relaxed ones on uniform clusters.}
  \label{fig:optimized_cluster}
\end{figure}

Note that the overlapping Algorithm \ref{alg:optimized_cluster} maybe not be very efficient in practice, as it may generate a large list of clusters. For example, for our toy model and $k = 2$, it would give a 
list of $20$ clusters. However, we expect from Figure \ref{fig:errordecay} that grouping the first few terms in optimized clusters, which correspond to dominant 
correlations $I(i,j)_{\rho^{\rm sdp}}$, suffices to produce satisfactory error reduction. To justify this, we sequentially group the clusters in the output list of 
Algorithm \ref{alg:optimized_cluster} for both overlapping and non-overlapping cases and plot the associated relaxed energies $E_0^{{\rm sdp}}/N$ in Figure 
\ref{fig:optimized_cluster}. For example, the third point in Figure \ref{fig:optimized_cluster} (right) corresponds to $E_0^{{\rm sdp}}/N$ on the cluster $\{ \{3,4\}, \{6,9\},  \{5,11\}, 1, 2, 7, 8, 10, 12\}$. We observe that for both cases, three or four $2 \times 1$ optimized clusters are sufficient to achieve the same (even tighter) 
relaxation as the one on uniform $2 \times 1$ clusters, and the further grouping does not help too much. 

We also apply Algorithm \ref{alg:optimized_cluster} with optimized $2 \times 1$ clusters as the initial cluster to generate optimized $4 \times 1$ clusters. We group only the first cluster in the output list, resulting in 
$\{ \{3,4, 6,9\},  \{5,11\}, \{2,7\}, \{10,12\},  \{1,8\}\}$. We find that the associated relaxation (seventh point in Figure \ref{fig:optimized_cluster} (right)) is already tighter than the one on uniform $3 \times 1$ clusters. To justify the equivalence between $I(i,j)_{\rho}$ and $I(i,j)_{\rho^{\rm sdp}}$ for the cluster selection, in Figure \ref{fig:optimized_cluster}, we benchmark
against the relaxations on clusters generated by Algorithm \ref{alg:optimized_cluster} with exact correlations $I(i,j)_{\rho}$, and we see that there is almost no difference between the corresponding results.

In summary, we find that the mutual information can serve as a cluster-selection indicator in the sense that grouping $V_i$ and $V_j$ with larger correlation $I(V_i, V_j)_\rho$ yields a tighter relaxation. We observe that the non-overlapping optimized cluster is typically more efficient than the overlapping one, and grouping the first few optimized clusters can sufficiently tighten the relaxation without incurring excessive computational costs. Therefore, we will mainly focus on the non-overlapping cluster in the subsequent discussion. Based on these insights, we propose Algorithm \ref{alg:sequence}, which progressively groups clusters with high correlations to produce an SDP hierarchy for approximating $E_0$.

\begin{algorithm}[!htbp]
\caption{Variational embedding with successively optimized clusters}
\begin{algorithmic}[1]
\Require Hamiltonian $H$; Initial cluster $[d] = \bigcup_i V_i$ with $V_i = \{i\}$; 
\Statex \q\q Maximal iteration number $\texttt{max}$; 
\Statex \q\q Maximal cluster size $k$ (if no cluster limit, set $k=0$);
\Statex \q\q Number of optimized clusters to be updated $n$. 
\Ensure An array of relaxed energies. 
\State Let the \texttt{current cluster} $\{V_1,V_2,\ldots\}$ be the initial cluster. 
\For{$s$ from $1$ to $\texttt{max} - 1$}
\State Solve $E_0^{ \rm sdp}[H]$ on \texttt{current cluster} with marginals $\{\rho_{ij}^{\rm sdp}\}_{i < j}$ on $V_{ij}: = V_i \bigcup V_j$. 
\State Compute $I(V_i,V_j)_{\rho^{\rm sdp}}$ for each $i < j$.
\State Order clusters $V_{ij}$ by $I(V_i,V_j)_{\rho^{\rm sdp}}$ as in \eqref{def:order_pair}. 
\State Define the ordered list by $\texttt{Pair} := \{V_{ij}\}$.
\If{$k > 0$}
\State Remove clusters of size larger than $k$ from $\texttt{Pair}$.
\EndIf
\State Define an empty list for new clusters $\texttt{List} = \{\}$.
\For{$l$ from $1$ to $n$}
\State Find the first $V_{ij}$ in $\texttt{Pair}$ such that $V_{ij} \bigcap\, (\text{ union of  clusters in $\texttt{List}$}) = \emptyset$. 
\State Add $V_{ij}$ into $\texttt{List}$ and remove it from $\texttt{Pair}$.  
\EndFor
\State Define an index set $\texttt{Index} = \bigcup\{i,j; \ V_{ij}\ \text{ in the \texttt{List}}\}$. 
\State Define the \texttt{current cluster} $= \bigcup\{V_{ij}\ \text{in the \texttt{List}}\} \bigcup \{V_i\ \text{with}\ i \notin \texttt{Index}\}$. 
\EndFor
\State Solve $E_0^{\rm sdp}[H]$ on the \texttt{current cluster}. 
\State Return the sequence of $E_0^{\rm sdp}$. 
\end{algorithmic} \label{alg:sequence}
\end{algorithm}

\begin{figure}[!htbp]
  \centering
  \includegraphics[width=0.45\textwidth]{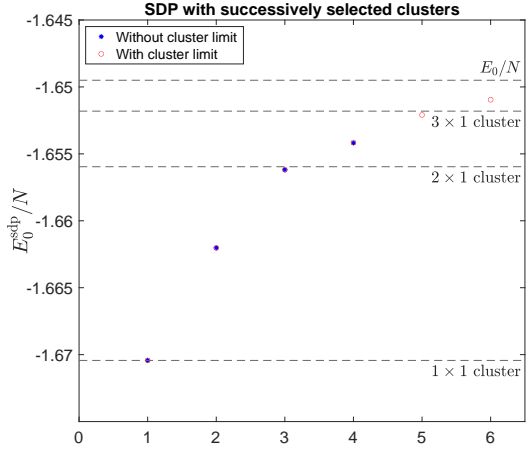}
  \caption{Successive SDP relaxations from Algorithm \ref{fig:optimalcluster1} with cluster size limit ($k = 3$, $\texttt{max} = 6$, and $n = 1$) and without size limit ($k = 0$, $\texttt{max} = 4$, and $n = 1$).}
  \label{fig:optimalcluster1}
\end{figure}

We test Algorithm \ref{alg:sequence} for the toy TFI model with the cluster size limit $k = 3$ and without the limit and set the number $n = 1$ of optimized clusters to be updated. The results are depicted in Figure \ref{fig:optimalcluster1}. We find that when there is no cluster size limit, 
Algorithm \ref{alg:sequence} is prone to generating increasingly large single clusters, which leads to the exponential growth of computational costs as the iteration proceeds. For instance, in our toy model, the 
fifth iteration produces a $5 \times 1$ cluster $\{3, 4, 6, 9, 11\}$, and the associated SDP is of size $O(10^6)$, which is prohibitively expensive to compute. However, as Figure \ref{fig:optimalcluster1} shows, with the cluster size constraint, Algorithm \ref{alg:sequence} can successfully generate a sequence of relaxed energies approximating $E_0$ from below.  It is worth noting that the cluster size limit $k$ comes with a caveat. 
The tightest relaxation in the output sequence of Algorithm \ref{alg:sequence}, which is of greatest interest, may not surpass the relaxation on the optimized $k \times 1$ clusters from Algorithm \ref{alg:optimized_cluster}
and, at the same time, have a larger computational cost, as it requires solving the SDP repeatedly. In practice, one must carefully choose the parameters
$\texttt{max}$ and $k$ in Algorithm \ref{alg:sequence} to balance the cost and accuracy.

\begin{figure}[!htbp]
  \centering
  \includegraphics[width=0.8\textwidth]{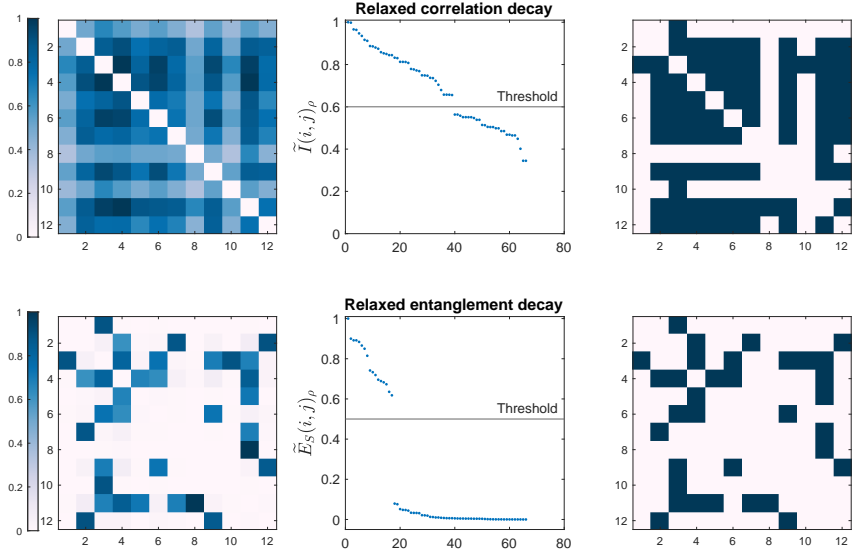}
  \caption{Normalized correlation and entanglement distributions $\w{I}(i,j)_{\rho^{\rm sdp}}$  and $\w{E}_S(i,j)_{\rho^{\rm sdp}}$ (first column) and their decay properties (second column), as well as the truncations \eqref{eq:truncation} by thresholds (third column).  
  }
  \label{fig:entanglement}
\end{figure}

We end this section with a few remarks. We have discussed the possibility of using the total 
correlation as the indicator to select optimal clusters in the SDP relaxations. We will see in 
Section \ref{sec:numerical} that the efficiency of such an approach strongly depends on the phase diagram of the given Hamiltonian. Thus, for a specific problem, a more efficient method could exist. We recall from Figure \ref{fig:errordecay} that the entanglement measure $E_S(A,B)_\rho$ may not be effective in optimizing the clusters. However, it is very interesting to observe that it could be used to recover the underlying graph of the Hamiltonian. In detail, we 
consider the normalized correlations and entanglements with relaxed marginals: for $i < j$, 
\begin{align} \label{eq:truncation}
    \w{I}(i,j)_{\rho^{\rm sdp}} = \frac{ I(i,j)_{\rho^{\rm sdp}}}{\max \{I(i,j)_{\rho^{\rm sdp}}\}}\,,\q   \w{E}_S(i,j)_{\rho^{\rm sdp}} = \frac{ E_S(i,j)_{\rho^{\rm sdp}}}{\max \{E_S(i,j)_{\rho^{\rm sdp}}\}}\,, 
\end{align}
which can be regarded as symmetric $d \times d$ matrices by setting $\w{I}(i,i)_{\rho^{\rm sdp}} = \w{E}_S(i,i)_{\rho^{\rm sdp}} = 0$ for $i \in [d]$. We visualize their magnitudes in Figure \ref{fig:entanglement} (first column). It shows that the entanglement can perfectly characterize the underlying graph $G$ in Figure \ref{fig:graph1}, namely, $\w{E}_S(i,j)_{\rho^{\rm sdp}}$ is large if and only if $(i,j)$ is an edge of $G$.  To be specific, we sort $\{\w{E}_S(i,j)_{\rho^{\rm sdp}}\}_{i < j}$ in decreasing order, denoted as $\{c_l\}$. We say that a gap exists between $c_l$ and $c_{l+1}$ if $l$ achieves the largest relative magnitude $\max_l \{c_l/c_{l+1}\}$. We pick an arbitrary threshold
value $c_l < p_0 < c_{l+1}$ and define the reconstructed adjacency matrix by the following logical matrix: 
\begin{align*}
    \w{E}^{\rm thres}_S(i,j) = \left\{ 
   \begin{aligned}
   &1, &&\text{if}\ \w{E}_S(i,j)_{\rho^{\rm sdp}} \ge p_0, \\
   &0, && \text{otherwise}.
   \end{aligned}
    \right.
\end{align*}
As shown in Figure \ref{fig:entanglement} (second row), $\w{E}^{\rm thres}_S(i,j)$ exactly recovers the adjacency matrix of $G$. We present the corresponding result to the total correlation in Figure 
\ref{fig:entanglement} (first row). We can see that although there is still an apparent gap for $\w{I}(i,j)_{\rho^{\rm sdp}}$, the associated logical matrix overestimates the adjacency matrix 
severely. Experiments in Section \ref{sec:numerical} show that the existence of such a gap in the entanglement with the perfect recovery of the underlying graph seems to be very robust for the 
ferromagnetic TFI model. It would be interesting to 
develop a mathematical theory to explain these observations. A simple corollary from Figures \ref{fig:errordecay} and \ref{fig:entanglement} is that grouping the edges of the graph $G$ is not an efficient way to tighten the relaxation.

\section{Numerical experiments} \label{sec:numerical}
This section aims at demonstrating the effectiveness of our variational embedding with optimized clusters (Algorithms \ref{alg:optimized_cluster} and \ref{alg:sequence}) for 
ground-state energy problems and the capability of utilizing  quantum entanglement to reconstruct the underlying graph of the Hamiltonian. We consider three prototypical quantum many-body problems: the TFI model \eqref{def:tfimodel}, the XXZ model \eqref{def:xxzmodel}, and the $t$--$U$ Hubbard model of spinless fermions 
\eqref{eq:hamiltonian_spinless}. All the experiments are conducted on a personal laptop using the MATLAB software package CVX for 
the SDP problems \cite{cvx} with the SDPT3 solver \cite{toh1999sdpt3}. We restrict the problem size to be small so that the exact diagonalization is applicable. 

\subsection{Transverse-field Ising model} 
\begin{figure}[!htbp]
    \centering
    \includegraphics[width=0.9\textwidth]{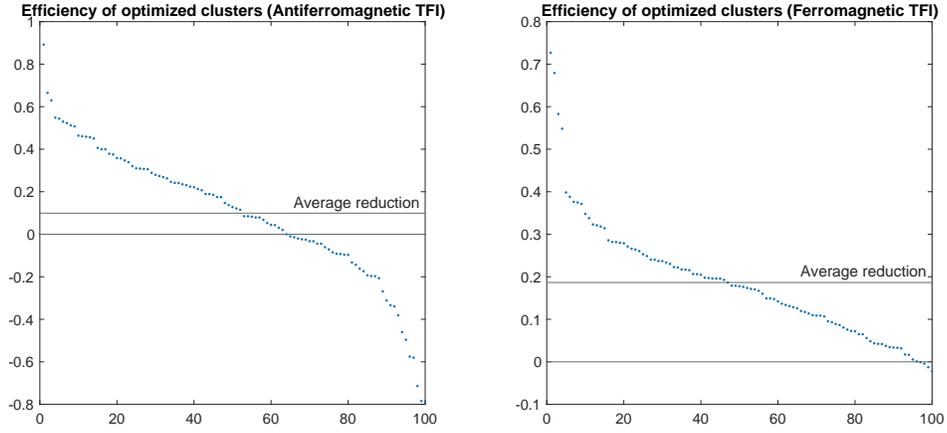}
    \caption{Efficiency (error reduction) factors ${\rm I}_{\rm eff}$ of optimized $2 \times 1$ clusters by Algorithm \ref{alg:optimized_cluster} for the TFI model with $J_{ij} = -1,\,h_i = 1$ (left) and $J_{ij} = 1,\,h_i = 1$ (right) on    
    100 random graphs generated by Erd\"{o}s–R\'{e}nyi model $G(8,0.4)$, in decreasing order.}
    \label{fig:tfi_random}
\end{figure}

We start with the TFI model \eqref{def:tfimodel}. In order to quantify the error reduction by cluster optimization, we consider the optimized $2 \times 1$ cluster selected by Algorithm \ref{alg:optimized_cluster} and define the efficiency factor by
\begin{align} \label{def:eff_fac}
    {\rm I}_{\rm eff} := \frac{E_0^{2\times 1, {\rm opt}}- E_0^{2\times 1, {\rm uni}}}{E_0- E_0^{2\times 1, {\rm uni}}}\,,
\end{align}
where $E_0^{2\times 1, {\rm uni}}$ and $E_0^{2\times 1, {\rm opt}}$   denote SDP relaxations on the uniform and optimized $2 \times 1$ clusters, respectively. We generate $100$ graph instances by Erd\"{o}s–R\'{e}nyi model $G(N,p)$ with $N = 8$ and $p = 0.4$, and compute the  associated factors ${\rm I}_{\rm eff}$ for both the ferromagnetic case: $J_{ij} = 1,\,h_i = 1$ and 
the antiferromagnetic case: $J_{ij} = -1,\,h_i = 1$ which is known to be generally  more difficult for the existence of frustration. We plot ${\rm I}_{\rm eff}$ in Figure \ref{fig:tfi_random} in decreasing order and observe that the antiferromagnetic case has a smaller mean $0.0993$ but with a larger variance, compared to the ferromagnetic one which has the average reduction $0.1865$ with variance $0.0192$. Moreover, Figure \ref{fig:tfi_random} shows that our strategy provides a tighter  relaxation for only $64 \%$ tested antiferromagnetic problems, while it is $96 \%$ in the ferromagnetic case. 

We next fix a graph instance generated by $G(12,0.4)$ shown in Figure \ref{fig:exp2graph} below. Similarly to Figure \ref{fig:errordecay}, for coefficients $J_{ij} = -1,\,h_i = 1$ and $J_{ij} = 1,\, h_i = 1$, we compute the relaxed energy $E_0^{{\rm sdp}}(i,j)$ on $1 \times 1$ clusters with the single $2 \times 1$ cluster $\{i,j\}$ and plot it in $(i,j)$ ordered by the total correlation \eqref{def:order_pair} in Figure \ref{fig:exp2transition}. We find that in the 
antiferromagnetic regime, the tendency for the error $(E_0 - E_0^{\rm sdp}(i,j))/N$ to increase as $I(i,j)_\rho$ decreases  is quite weak. From these observations, we can conclude that our cluster-selection strategy in the antiferromagnetic case is not as efficient as in the ferromagnetic case.

\begin{figure}[!htbp]
    \centering
    \includegraphics[width=0.65\textwidth]{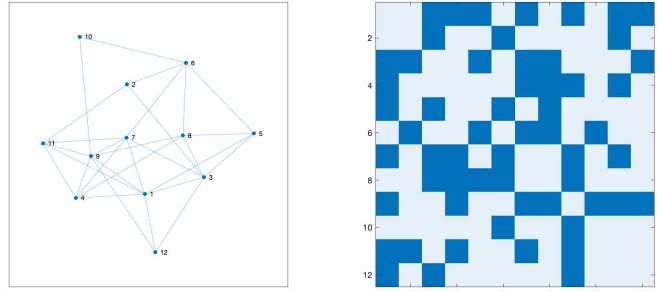}
    \caption{An instance of $G(12,0.4)$ (left) and its adjacency matrix (right) for the TFI model.}
    \label{fig:exp2graph}
\end{figure}

\begin{figure}[!htbp]
    \centering
    \includegraphics[width=0.9\textwidth]{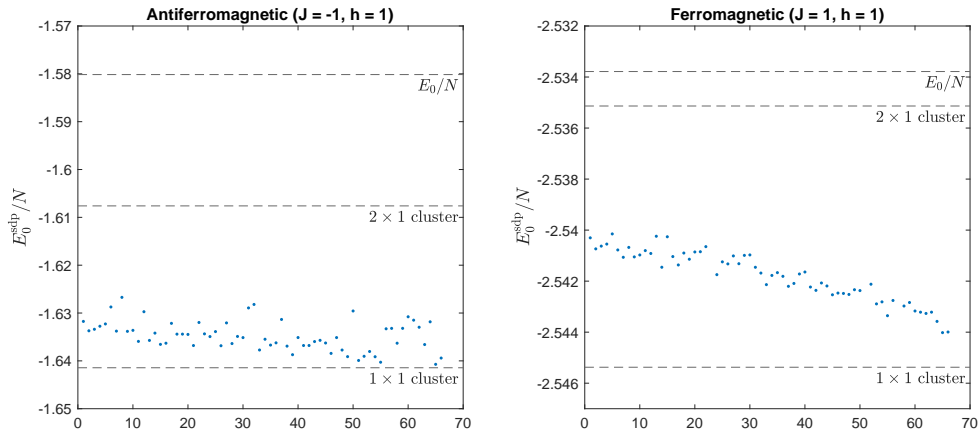}
    \caption{Relaxed energy per site $E_0^{\rm sdp}(i,j)/N$ in $\{(i,j)\}_{i<j}$ ordered by the total correlation with exact marginals for the antiferromagnetic (left) and ferromagnetic (right) TFI models on the graph shown in Figure \ref{fig:exp2graph}. 
    }
    \label{fig:exp2transition}
\end{figure}

\begin{figure}[!htbp]
    \centering
    \includegraphics[width=0.45\textwidth]{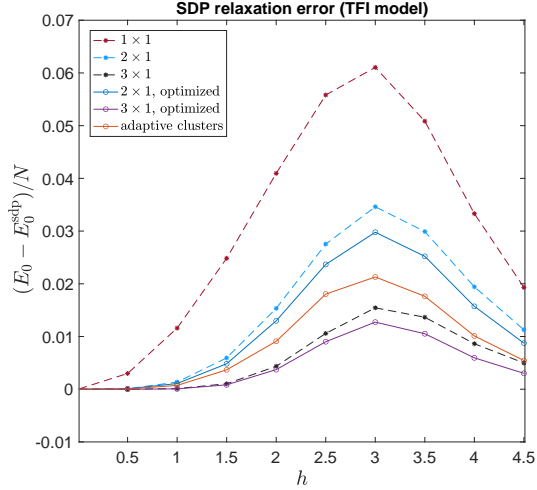}
    \caption{SDP relaxation errors for optimized clusters generated by Algorithms \ref{alg:optimized_cluster} and \ref{alg:sequence} for the TFI model on the graph in Figure \ref{fig:exp2graph} with coefficients $J_{ij} = 1$ and various $h$.
    }
    \label{fig:tfierror}
\end{figure}

Now, let us consider the ferromagnetic TFI model on the graph in Figure \ref{fig:exp2graph} with $J_{ij} = 1$ and various $h$. We plot in  
Figure \ref{fig:tfierror} the SDP relaxation errors $(E_0 - E_0^{\rm sdp})/N$ for optimized $2 \times 1$ and $3 \times 1$ 
clusters generated by Algorithm \ref{alg:optimized_cluster}. We also test the successively selected clusters by Algorithm \ref{alg:sequence}, where we set the iteration number $\texttt{max} = 5$ and cluster size limit $k= 3$ so that the total computational time is approximately half that of the SDP on uniform $3 \times 1$ clusters. The cluster yielding the tightest relaxation in the output sequence of Algorithm  \ref{alg:sequence} is referred to as the adaptive cluster; the associated relaxation errors are plotted in Figure \ref{fig:tfierror}.
In our tests, the adaptive cluster consists of $1 \times 1$ clusters along with a $3 \times 1$ cluster and two $2 \times 1$ clusters, or two $3 \times 1$ clusters. To provide a benchmark, we compare these results against the errors for uniform $1 \times 1$, $2 \times 1$, and $3 \times 1$ clusters. It is evident from Figure \ref{fig:tfierror} that the optimized clusters result in significant error reductions compared to the uniform ones.

\begin{figure}[htbp]
    \centering
    \includegraphics[width=0.9\textwidth]{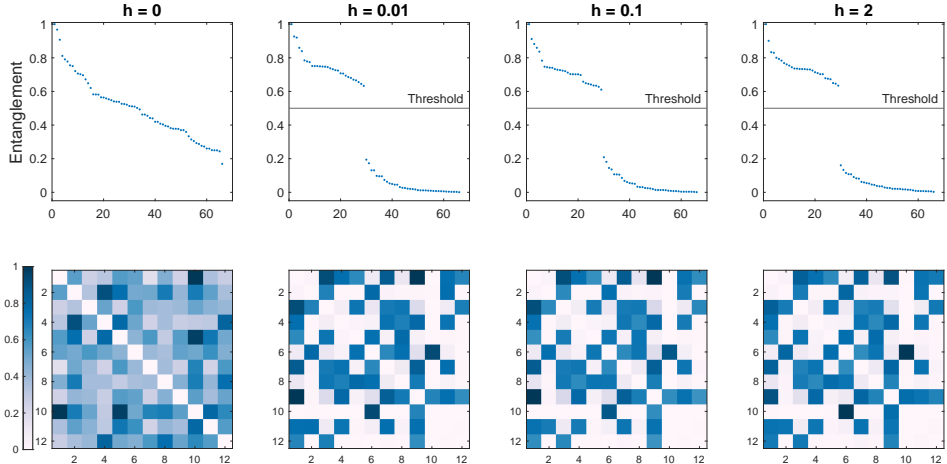}
    \caption{Normalized entanglement distributions $\w{E}_S(i,j)_{\rho^{\rm sdp}}$ (second row) and their decay properties (first row) for the TFI model on the graph in Figure \ref{fig:exp2graph} with $J_{ij} = 1$ and various $h$.}
    \label{fig:exp2entangle}
\end{figure}

In order to further explore the reconstruction of the underlying graph through quantum entanglement, we compute the normalized entanglements
$\w{E}_S(i,j)_{\rho^{\rm sdp}}$ for various values of $h$, and plot their decaying patterns and 
distributions in Figure \ref{fig:exp2entangle}. Notably, we observe a sharp transition in the gap existence at $h = 0$. 
In other words, the entanglement distribution for $h = 0$ (which corresponds to the classical ferromagnetic Ising model) fails to accurately recover the graph structure of the Hamiltonian, while 
a very small transverse field suffices to give the exact and robust reconstruction, as shown in Figure \ref{fig:exp2entangle} (second row). 

\begin{figure}[htbp]
    \centering
    \includegraphics[width=0.8\textwidth]{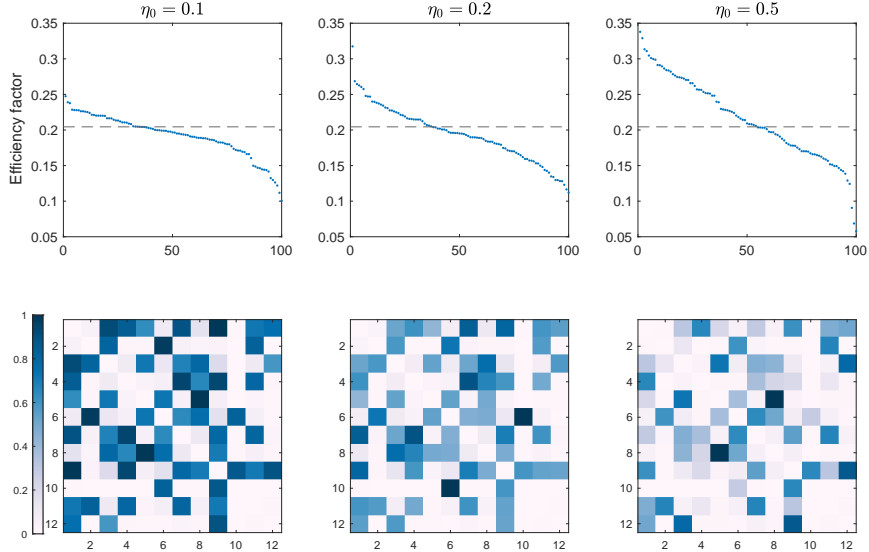}
    \caption{First row: efficiency factors ${\rm I}_{\rm eff}$ of optimized $2 \times 1$ clusters for 100 instances of the disordered TFI model with $h_i = 1$ and $J_{ij} \sim \mc{N}(1,\eta_0^2)$ with $\eta_0 = 0.1$ (average ${\rm I}_{\rm eff} = 0.1916$, variance $8.65\times10^{-4}$), $\eta_0 = 0.2$ (average ${\rm I}_{\rm eff} = 0.1939$, variance $0.0015$), and $\eta_0 = 0.5$ (average ${\rm I}_{\rm eff} = 0.2154$, variance $0.0033$), in decreasing order.     Second row: normalized entanglement distributions $\w{E}_S(i,j)_{\rho^{\rm sdp}}$ for one of the 100 instances with $\eta = 0.1, 0.2, 0.5$ from left to right.}
    \label{fig:tfi_disorder}
\end{figure}

Finally, we investigate the robustness of our strategy against the noise by considering the disordered TFI model on the graph in Figure \ref{fig:exp2graph} with $h_i = 1$ and $J_{ij}$ being Gaussian $\mc{N}(1,\eta_0^2)$. In Figure \ref{fig:tfi_disorder} (first row), we compute the efficiency factors \eqref{def:eff_fac} for $100$ instances of this disordered TFI model with $\eta_0 = 0.1, 0.2, 0.5$, respectively. The dashed line ${\rm I}_{\rm eff} = 0.2044$ represents the efficiency factor for $J_{ij} = 1$ (the case without disorder). We observe that the average error reduction is stable with respect to the disorder strength $\eta_0$, but the variance increases as $\eta_0$ increases, as one can expected. In Figure \ref{fig:tfi_disorder} (second row), we display the normalized entanglement distributions $\w{E}_S(i,j)_{\rho^{\rm sdp}}$ associated with one of the instances for $\eta_0 = 0.1, 0.2, 0.5$, respectively.  We find that the entanglement pattern is robust against the relatively small disorder, and can still recover the underlying graph structure accurately.

\subsection{XXZ model} We next consider the spin-$1/2$ XXZ model \eqref{def:xxzmodel}. We first plot in Figure 
\ref{fig:xxz_sta} the efficiency factor that quantifies the error reduction by optimized clusters in decreasing order, for 
$100$ graph instances from Erd\"{o}s–R\'{e}nyi model $G(8,0.4)$. 
We find that our approach can efficiently tighten the SDP relaxations for $94\%$ of the tested problems, achieving an average error reduction of $41.11\%$ with a variance of $0.0607$.

\begin{figure}[!htbp]
    \centering
    \includegraphics[width=0.65\textwidth]{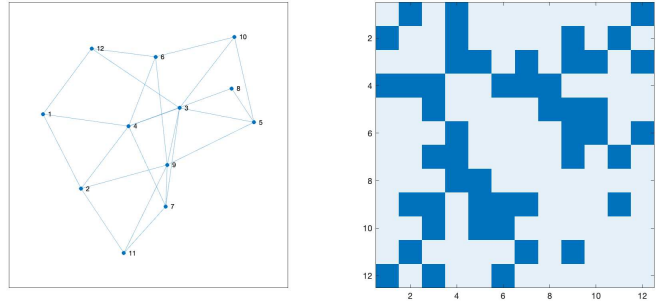}
    \caption{An instance of $G(12,0.3)$ (left) and its adjacency matrix (right) for XXZ model.}
    \label{fig:xxz_graph}
\end{figure}

\begin{figure}[!htbp]
    \centering
     \begin{subfigure}[b]{0.48\textwidth}
    \includegraphics[width=\linewidth]{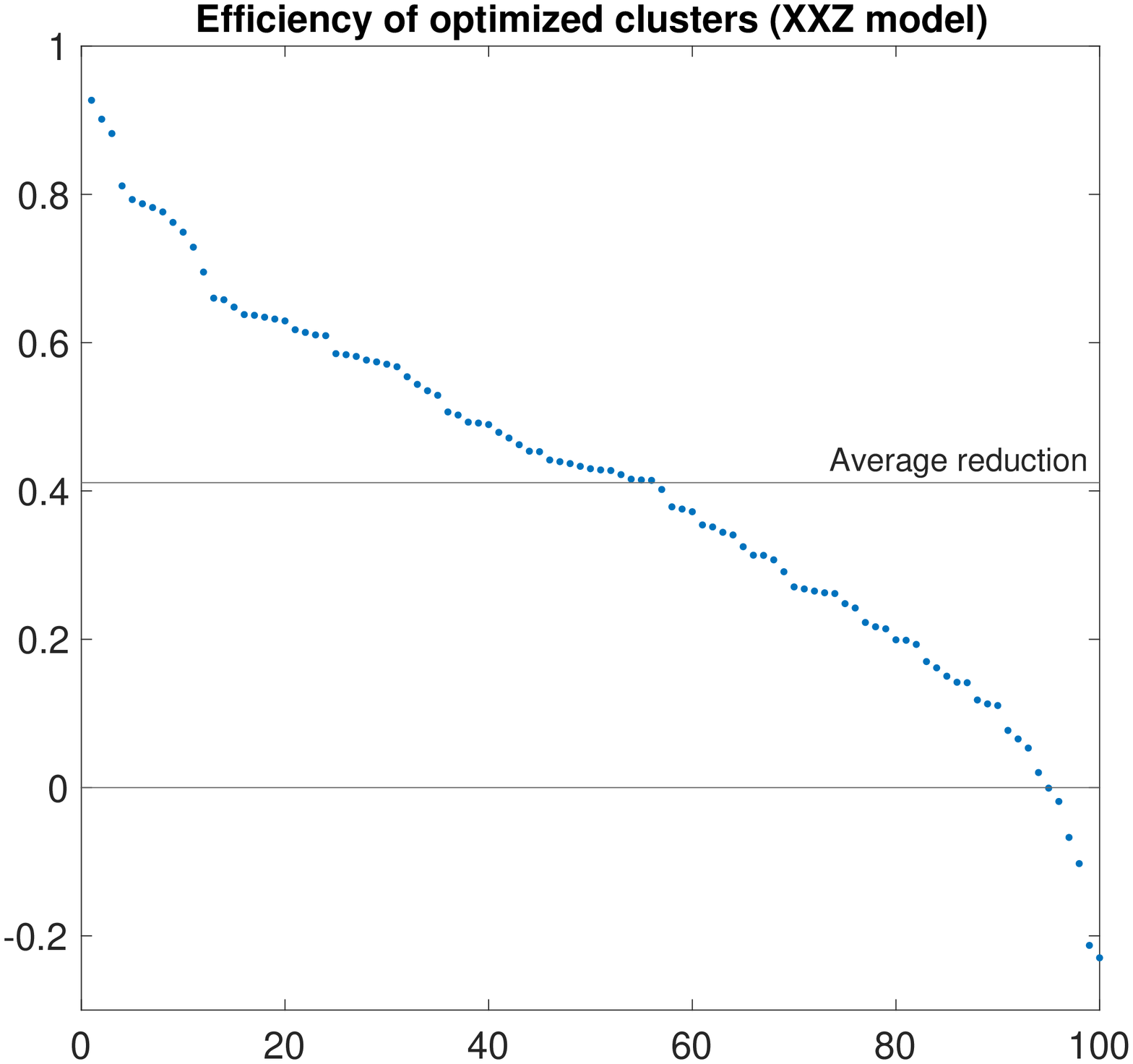}
     \subcaption[]{}
     \label{fig:xxz_sta}
\end{subfigure}
   \begin{subfigure}[b]{0.48\textwidth}
    \includegraphics[width=\linewidth]{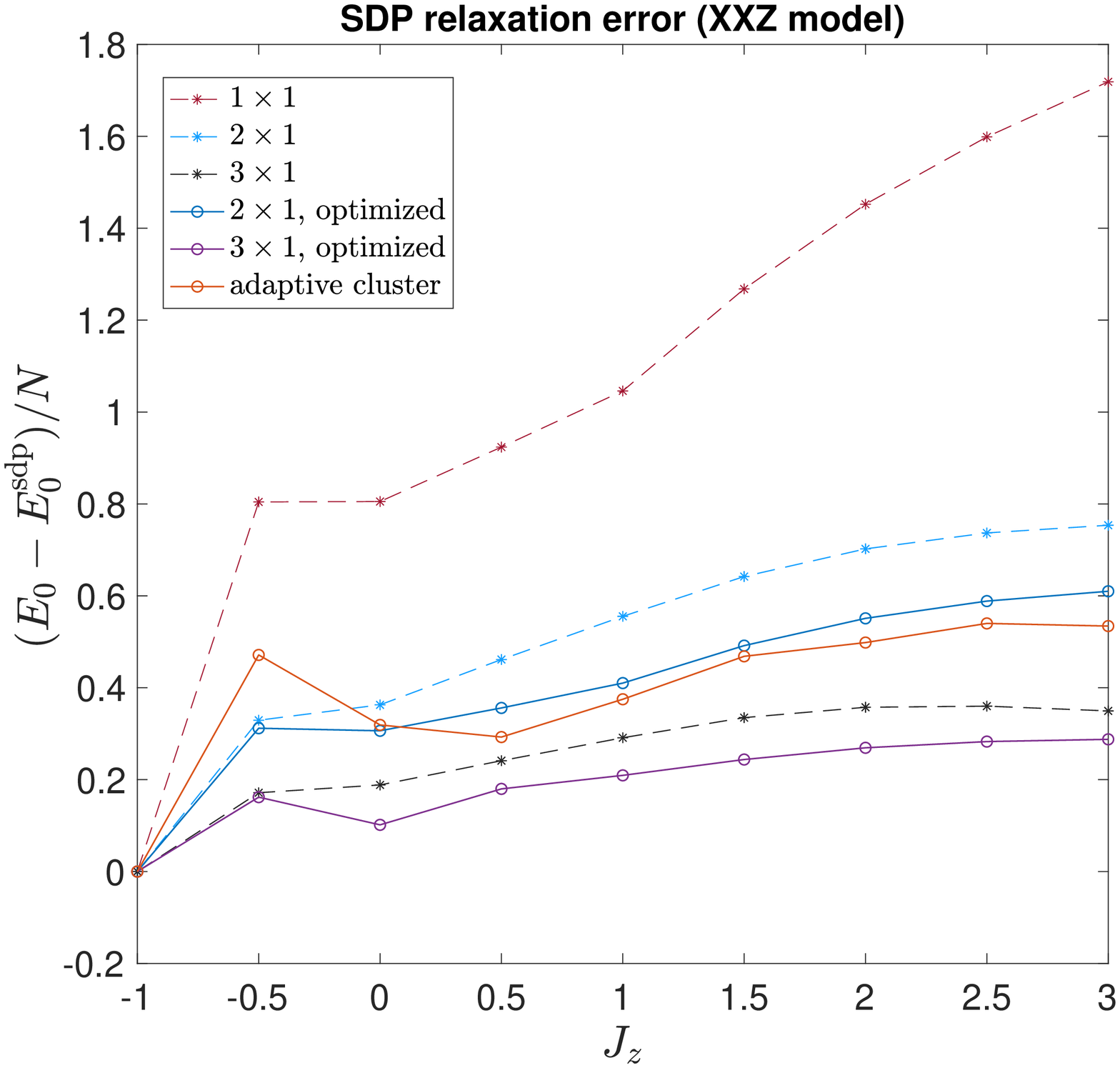}
    \subcaption[]{}
   \label{fig:xxz_error1}
\end{subfigure}
\caption[]{Results for the XXZ model: (a) efficiency factors ${\rm I}_{\rm eff}$ of optimized $2 \times 1$ clusters for $J_z = 1$ and $100$ random graphs generated by $G(8,0.4)$, in decreasing order. (b) SDP relaxation errors for optimized clusters generated by Algorithms \ref{alg:optimized_cluster} and \ref{alg:sequence} with the underlying graph in Figure \ref{fig:xxz_graph} with various $J_z$.}
\end{figure}

We now fix a random graph generated by the model $G(12,0.3)$ shown in Figure \ref{fig:xxz_graph}, in order to test the effectiveness of Algorithms \ref{alg:optimized_cluster} and \ref{alg:sequence} on the XXZ model with various $J_z$. Similarly to the experiments for TFI model, in Figure \ref{fig:xxz_error1}, we compute the relaxation errors $(E_0 - E_0^{\rm sdp})/N$ for $2 \times 1$ and $3 \times 1$ optimized clusters given by Algorithm \ref{alg:optimized_cluster} and the adaptive cluster generated by Algorithm \ref{alg:sequence} with $\texttt{max} = 5$ and $k = 3$, and benchmark the results against the ones on uniform $1 \times 1$, $2 \times 1$, and $3 \times 1$ clusters. Note that all the relaxations are exact at $J_z = -1$. Moreover, there is a transition of the effectiveness of optimized clusters at $J_z = 0$. When $J_z < 0$, the improvements by optimizing the cluster are limited or even worse (cf. adaptive clusters from Algorithm~\ref{alg:sequence}). However, when $J_z > 0$, we can clearly observe that the optimized cluster can achieve the apparent error reductions, given the same computational resource.

\begin{figure}[!htbp]
    \centering
    \includegraphics[width=0.8\textwidth]{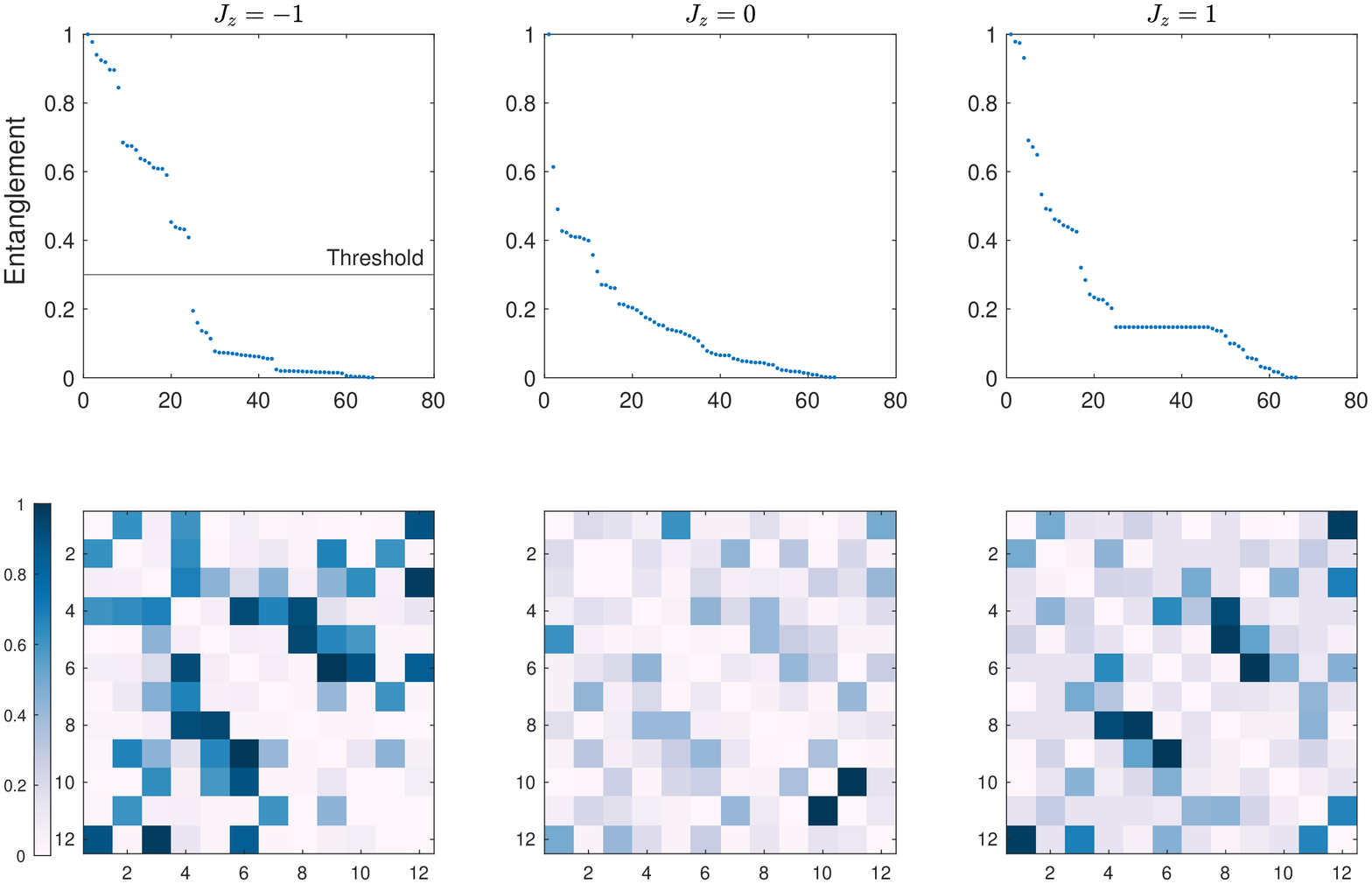}
    \caption{Normalized entanglement distributions $\w{E}_S(i,j)_{\rho^{\rm sdp}}$ (second row) and their decay properties (first row) for the XXZ model on the graph in Figure \ref{fig:xxz_graph} with various $J_z$.}
    \label{fig:xxz_ent}
\end{figure}

We discuss the possibility of using the entanglement distribution to extract the underlying graph information of the XXZ model. We plot the normalized entanglements 
$\w{E}_S(i,j)_{\rho^{\rm sdp}}$ and their decaying patterns for various $J_z$ in Figure \ref{fig:xxz_ent}. It turns out that for the XXZ model, the quantum entanglement may not be as efficient as it is 
for the TFI model. Comparing Figures \ref{fig:xxz_graph} and \ref{fig:xxz_ent}, we see that only in the case of $J_z = -1$, the gap exists and allows us to reconstruct the graph structure by entanglement. 

\subsection{Spinless Hubbard model}
\begin{figure}
    \centering
    \includegraphics[width=0.65\textwidth]{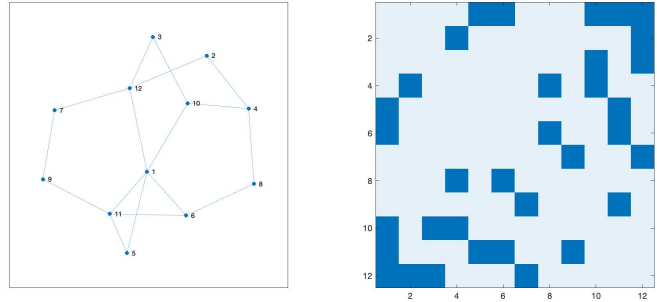}
    \caption{An instance of $G(12,0.3)$ (left) and its adjacency matrix (right) for the spinless Hubbard model.}
    \label{fig:hubbard_gra}
\end{figure}

\begin{figure}[!htbp]
    \centering
     \begin{subfigure}[b]{0.48\textwidth}
    \includegraphics[width=\linewidth]{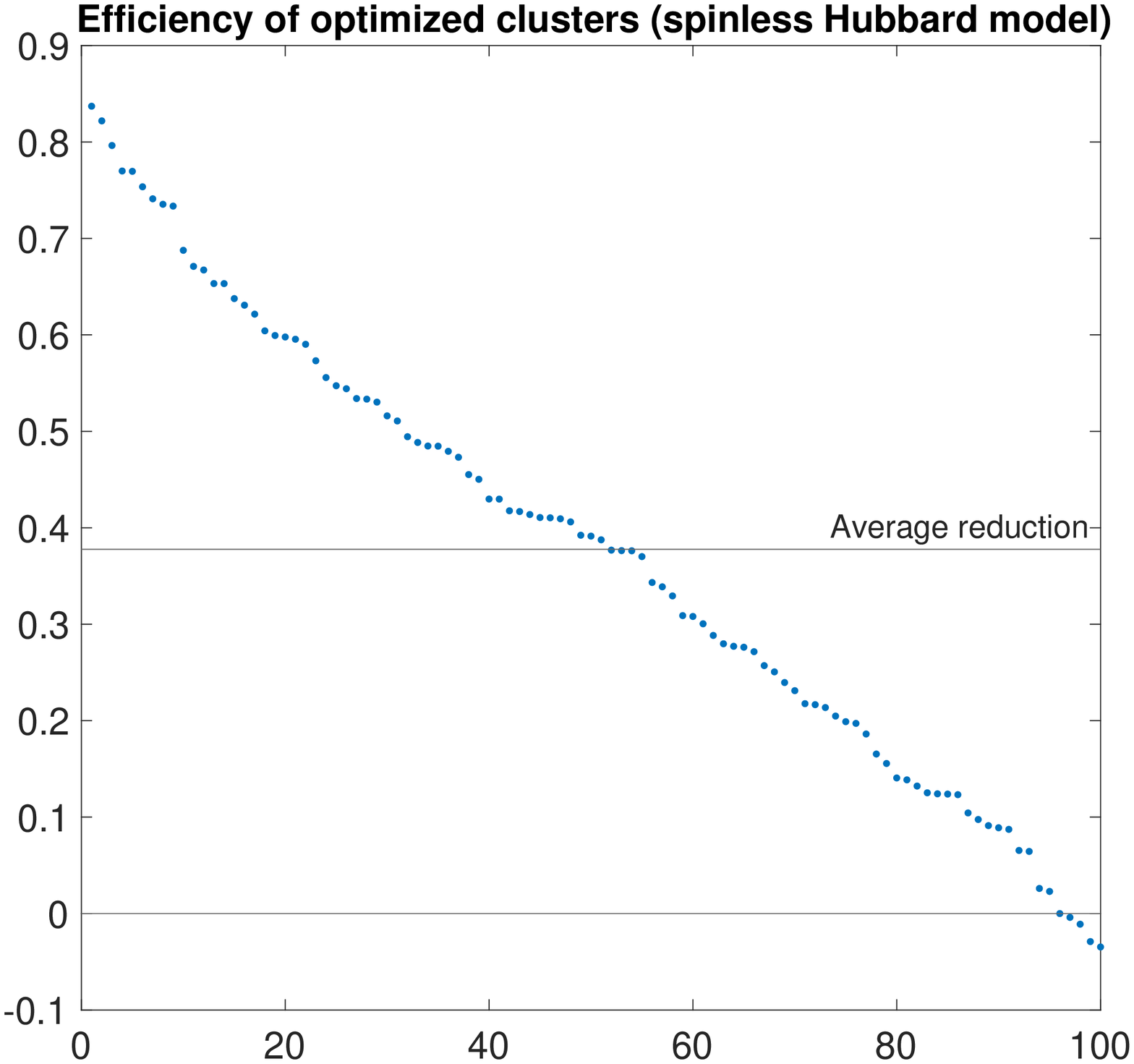}
     \subcaption[]{}
    \label{fig:hubbard_sta}
\end{subfigure}
   \begin{subfigure}[b]{0.48\textwidth}
    \includegraphics[width=\linewidth]{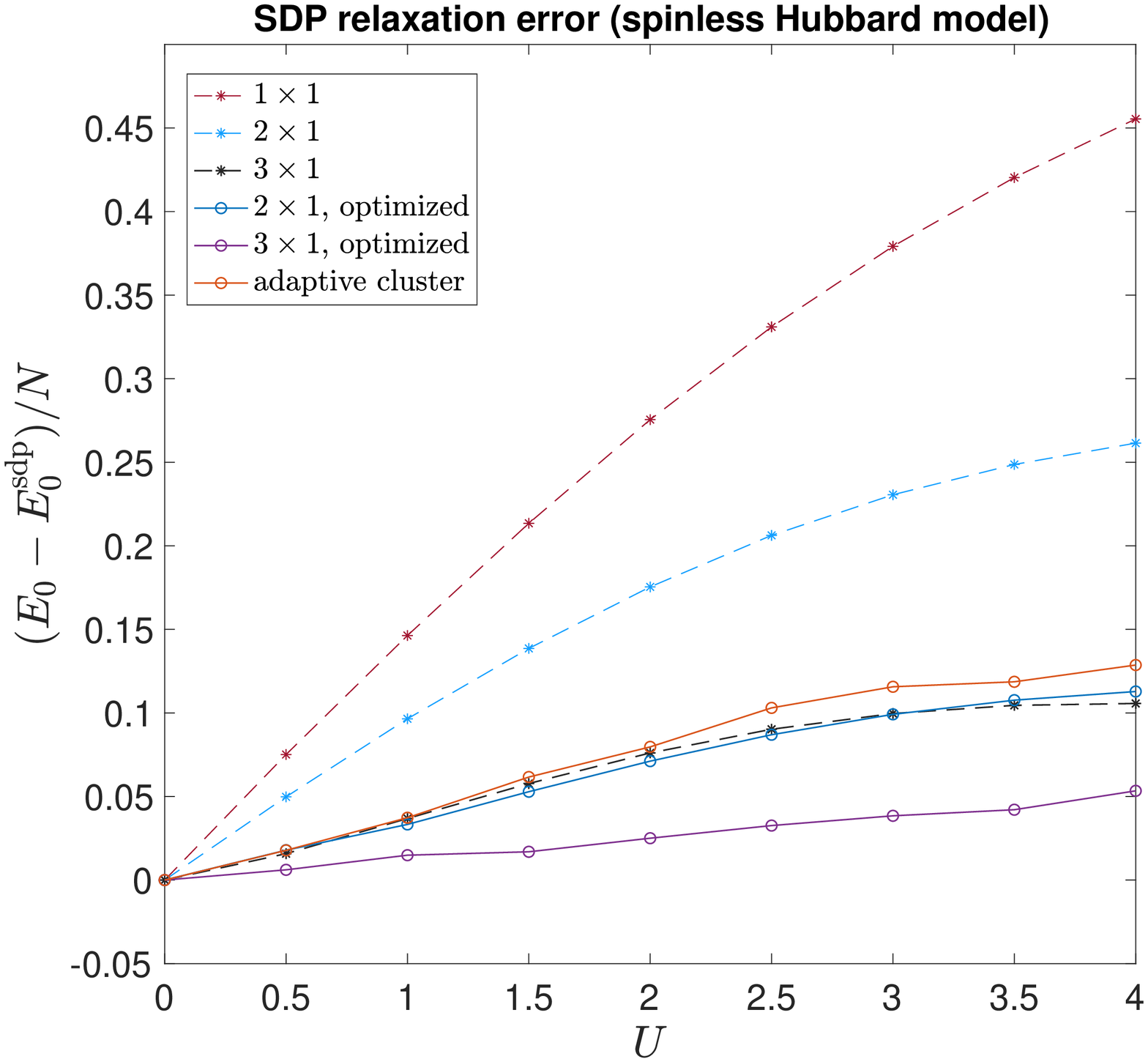}
    \subcaption[]{}
    \label{fig:hubbar_error}
\end{subfigure}
\caption[]{Results for the spinless Hubbard model: (a) efficiency factors ${\rm I}_{\rm eff}$ of optimized $2 \times 1$ clusters for $U = 1$ and $100$ random graphs generated by $G(8,0.4)$, in decreasing order. (b) SDP relaxation errors for optimized clusters generated by Algorithms \ref{alg:optimized_cluster} and \ref{alg:sequence} with the underlying graph in Figure \ref{fig:xxz_graph} and various $U$.}
\end{figure}

We finally consider the Hubbard model of spinless fermions \eqref{eq:hamiltonian_spinless}. Again, we first generate $100$ instances from Erd\"{o}s–R\'{e}nyi model $G(8,0.4)$ and compute associated efficient factors ${\rm I}_{\rm eff}$ to test the general error reduction by optimized $2 \times 1$ clusters in Figure \ref{fig:hubbard_sta}. Similarly, the optimized cluster can tighten the SDP relaxation for $95\%$ tested examples and achieve an error reduction of $37.93\%$ on average with a variance of $0.0502$. We then consider the model on a fixed graph instance shown in Figure \ref{fig:hubbard_gra}. We present the relaxations errors $(E_0 - E_0^{\rm sdp})/N$ for optimized $2 \times 1$ and $3 \times 1$ clusters by Algorithm \ref{alg:optimized_cluster} and the adaptive cluster by Algorithm \ref{alg:sequence} with benchmarking results on the uniform clusters. We clearly observe that the clusters selected by Algorithms \ref{alg:optimized_cluster} and \ref{alg:sequence} can help significantly reduce the relaxation errors. In particular, the errors for the optimized $2 \times 1$ cluster by Algorithm \ref{alg:optimized_cluster} and the adaptive one by Algorithm \ref{alg:sequence} can achieve almost the same error as the one for uniform $3 \times 1$ clusters while with much less computational costs.

\section{Concluding remarks}
In this work, we have generalized the variational 
embedding method for solving the ground-state energy problems from the sum-of-squares SDP hierarchy. We have detailed its connections with the RDM method and 
Anderson bounds. Moreover, we have discussed the possibility of using RDM conditions to tighten the 
variational embedding. Considering the inherent exponential scaling with the cluster size, we have proposed efficient strategies for optimizing the 
clusters to reduce the relaxation error, with given computational resources. As a byproduct, we have observed that quantum entanglement can help to recover the underlying 
graph of the many-body Hamiltonian.  
We conclude with several future directions.  Motivated by \cite{khoo2019convex}, it would be interesting to design a scheme that projects the relaxed marginals from variational embedding to the unrelaxed one~\eqref{eq:joint_rep}, 
which gives an upper bound to the ground-state energy. In addition, as discussed in Remark~\ref{rem:sublevel}, to overcome the exponential scaling in the cluster size, 
it is promising to develop a multilevel variational embedding method. Another challenging question is to find a mathematical illustration for the sharp entanglement transition observed in Figure~\ref{fig:exp2entangle}.

\end{document}